%% file: paper.tex
\documentclass[sigconf, 9pt]{acmart}
\usepackage[english]{babel}

\usepackage{siunitx}
\usepackage{amsmath,amsthm,amsfonts}
\usepackage{color}
\usepackage{graphicx}
\usepackage{subcaption}
\usepackage{xcolor}
\usepackage{hyperref}
\usepackage[nosort, nameinlink,capitalise]{cleveref}
\usepackage{makecell}
\usepackage{multirow}
\usepackage{booktabs}
\usepackage{tikz}
\usepackage{stmaryrd} 
\usepackage{mathtools}
\usepackage{caption}
\usepackage{subcaption}
\usepackage[nodisplayskipstretch]{setspace}
\usepackage{caption}
\usepackage{tikz}
\usepackage{soul}
\setstcolor{red}
\usepackage[shortlabels]{enumitem}

\DeclareSIUnit{\kB}{\kibi\byte}
\DeclareSIUnit{\MB}{\mebi\byte}
\DeclareSIUnit{\GB}{\gibi\byte}
\DeclareSIUnit{\TB}{\tebi\byte}
\DeclareSIUnit{\bps}{bps}
\DeclareSIUnit{\kbps}{\kilo\bps}
\DeclareSIUnit{\Mbps}{\mega\bps}
\DeclareSIUnit{\Gbps}{\giga\bps}
\DeclareSIUnit{\Tbps}{\tera\bps}
\DeclareSIUnit{\pps}{pps}
\DeclareSIUnit{\kpps}{\kilo\pps}
\DeclareSIUnit{\Mpps}{\mega\pps}
\DeclareSIUnit{\Gpps}{\giga\pps}
\DeclareSIUnit{\Tpps}{\tera\pps}
\DeclareSIUnit[number-unit-product = ]{\percent}{\char`\%} 
\DeclareSIUnit{\ton}{t}
\DeclareSIUnit{\year}{yr}
\DeclareSIUnit{\Gt}{\giga\ton}
\DeclareSIUnit{\Gtpa}{\Gt\per\year} 
\DeclareSIUnit{\tWh}{TWh}

\crefformat{section}{\S#2#1#3} 
\crefformat{subsection}{\S#2#1#3}
\crefformat{subsubsection}{\S#2#1#3}

\crefformat{equation}{Eq.~(#2#1#3)}
\crefformat{figure}{Fig.~#2#1#3}
\crefrangeformat{equation}{Eqs. #3(#1)#4--#5(#2)#6}

\newcommand{\ict}{ICT}

\newcommand{\co}{CO$_{2}$}

\newcommand{\copath}{carbon-aware path selection}
\newcommand{\cidt}{CIDT}

\newcommand{\systemname}{CIRo}
\newcommand{\ifpair}{\textit{[ing, eg]}}
\newcommand{\ifpairforward}{\textit{[ing, eg]}}
\newcommand{\ifpairbackward}{\textit{[eg, ing]}}
\newcommand{\fullpath}{\textit{\{AS}^{\textit{src}},\textit{ ...},\textit{ AS}^{\textit{dst}}\textit{\}}}
\newcommand{\ashop}{\textit{AS}_{\ifpair{}}}
\newcommand{\ashopforward}{\textit{AS}_{\ifpairforward{}}}
\newcommand{\ashopbackward}{\textit{AS}_{\ifpairbackward{}}}
\newcommand{\ashopi}{\textit{AS}^{i}_{\ifpair{}}}
\newcommand{\ciextension}{\texttt{CIDTEx\-ten\-sion}}
\newcommand{\staticextension}{\texttt{Stat\-icInfoEx\-ten\-sion}}
\newcommand{\asentry}{\texttt{ASEn\-try}}
\renewcommand{\paragraph}[1]{\noindent\textbf{#1.}}

\usepackage{pifont}
\newif\ifremovecomments
\removecommentsfalse

\makeatletter
\setcopyright{none}
\newcommand{\mysubscript}[2]{\textsubscript{\textcolor{#2}{\textsf{\textbf{#1}}}}}
\newcommand*\defcomment[4]{
	\ifremovecomments
	\expandafter\newcommand\csname #1\endcsname[1]{%
	}
	\expandafter\newcommand\csname @#2delnoname\endcsname[1]{%
	}
	\expandafter\newcommand\csname #2del\endcsname[1]{%
	}
	\expandafter\newcommand\csname #2sugg\endcsname[1]{##1}
	\expandafter\newcommand\csname #2subs\endcsname[2]{##2}
	\else
	\expandafter\newcommand\csname #1\endcsname[1]{%
		\textcolor{#4}{\ding{110}\mysubscript{#3}{#4}\,{##1}}%
	}
	\expandafter\newcommand\csname @#2delnoname\endcsname[1]{%
		\bgroup\markoverwith{\textcolor{#4}{\rule[0.35ex]{2pt}{1pt}}}\ULon{##1}%
	}
	\expandafter\newcommand\csname #2del\endcsname[1]{%
		\csname @#2delnoname\endcsname{##1}\kern0.1em\mysubscript{#3}{#4}%
	}
	\expandafter\newcommand\csname #2sugg\endcsname[1]{%
		\textcolor{#4}{[##1]\mysubscript{#3}{#4}}%
	}
	\expandafter\newcommand\csname #2subs\endcsname[2]{%
		\csname @#2delnoname\endcsname{##1}\csname #2sugg\endcsname{##2}%
	}
	\fi
	\expandafter\newcommand\csname #2sout\endcsname{\csname #2del\endcsname}
}

\renewcommand\footnotetextcopyrightpermission[1]{} 

\makeatother

\ifdefined\final
\newcommand{\sout}[1]{}

\newcommand{\todo}[1]{}
\newcommand{\mynote}[1]{}
\newcommand{\ml}[1]{}

\newcommand{\todoeurosnp}[1]{}
\newcommand{\adrian}[1]{}
\newcommand{\ali}[1]{}
\newcommand{\jk}[1]{}
\newcommand{\simon}[1]{}
\else
\usepackage[normalem]{ulem}

\newcommand{\todo}[1]{\textcolor{red}{\emph{\textbf{[TODO: }#1\textbf{]}}}}
\newcommand{\mynote}[1]{\textcolor{blue}{\emph{\textbf{NOTE: }#1}}}
\newcommand{\ml}[1]{\textcolor{green}{\emph{\textbf{NOTE (ML): }#1}}}

\newcommand{\todoeurosnp}[1]{\textcolor{orange}{\emph{\textbf{[TODO: }#1\textbf{]}}}}
\newcommand{\adrian}[1]{\textcolor{blue}{\emph{\textbf{AP: }#1}}}
\newcommand{\ali}[1]{\textcolor{brown}{\emph{\textbf{AT: }#1}}}
\newcommand{\simon}[1]{\textcolor{green}{\emph{\textbf{SS: }#1}}}
\newcommand{\jk}[1]{\hl{#1}}
\fi

\begin{document}

\title{Carbon-Intelligent Global Routing in Path-Aware Networks}

\author{Seyedali Tabaeiaghdaei}
\affiliation{%
  \institution{ETH Z\"urich}
  \city{Z\"urich} 
  \country{Switzerland}
}
\email{seyedali.tabaeiaghdaei@inf.ethz.ch}
\author{Simon Scherrer}
\affiliation{%
  \institution{ETH Z\"urich}
  \city{Z\"urich} 
  \country{Switzerland}
}
\email{simon.scherrer@inf.ethz.ch}
\author{Jonghoon Kwon}
\affiliation{%
  \institution{ETH Z\"urich}
  \city{Z\"urich} 
  \country{Switzerland}
}
\email{jonghoon.kwon@inf.ethz.ch}
\settopmatter{authorsperrow=2}
\author{Adrian Perrig}
\affiliation{%
  \institution{\mbox{ETH Z\"urich}}
  \city{Z\"urich} 
  \country{Switzerland}
}
\email{adrian.perrig@inf.ethz.ch}

\renewcommand{\shortauthors}{Tabaeiaghdaei et al.}

\input{abstract.tex}
\maketitle

\input{introduction}

\input{background}

\input{principles}
\input{model.tex}

\input{detailed_design.tex}

\input{global_impact/global_impact}

\input{discussion}
\input{related}
\input{conclusion}

\section*{ACKNOWLEDGEMENTS}

We would like to thank Markus Legner, Cedric Meury, and Tobias Schmid for their insightful discussions and valuable ideas and comments.

\label{end:paper}

\bibliographystyle{ACM-Reference-Format}
\bibliography{reference}

\appendix
\input{scion_specific_path_combination.tex}
\input{traffic_matrix.tex}

\input{related_models.tex}
\input{numbers_tables.tex}

\label{end:references}
\end{document}

%% file: abstract.tex
\begin{abstract}
The growing energy consumption of Information and Communication Technology (ICT) has raised concerns about its environmental impact. However, the carbon efficiency of data transmission over the Internet has so far received little attention. This carbon efficiency can be enhanced effectively by sending traffic over car\-bon-ef\-fi\-cient inter-domain paths. However, challenges in estimating and disseminating carbon intensity of inter-domain paths have prevented carbon-aware path selection from becoming a reality.

In this paper, we take advantage of path-aware network architectures to overcome these challenges. In particular, we design~\systemname{}, a system for forecasting the carbon intensity of inter-domain paths and disseminating them across the Internet. We implement a proof of concept for~\systemname{} on the codebase of the SCION path-aware Internet architecture and test it on the SCIONLab global research testbed.
Further, we demonstrate the potential of~\systemname{} for reducing the carbon footprint of endpoints and end domains through large-scale simulations. We show that~\systemname{} can reduce the carbon intensity of communications by at least 47\% for half of the domain pairs and the carbon footprint of Internet usage by at least 50\% for 87\% of end domains.

\end{abstract}

%% file: introduction.tex
\section{Introduction}
\label{sec:intro}
%

In the face of growing concerns regarding climate change, companies are under increasing pressure to measure and reduce their carbon footprint. This pressure also applies to their use of Information and Communication Technology (\ict{}) as ICT has a notable contribution of 2.7\% to global \co{} emissions~\cite{Lorincz2019co2estimate}, which is expected to grow significantly---approximately four times---until 2030~\cite{andrae2015trendsto2030}. Hence, reducing the carbon footprint of \ict{} use is becoming increasingly relevant for enterprises,  manifesting in carbon-neutrality statements of major technology corporations. For example, Google and Facebook already use 100\% renewable energy for their operations~\cite{google-cloud-sustainability-2023,guardian-facebook-sustainability-2021}.

While these efforts are laudable and impactful, promising opportunities for further carbon-footprint reduction exist. Indeed, previous research has identified a range of such opportunities. However, most of these proposals apply to local aspects: intra-domain networking (i.e., within a single domain), data-center optimizations, or neighbor-domain cooperation (cf.~\cref{sec:related}). In contrast, inter-domain networking (i.e., among multiple domains), which accounts for around 13\% of total ICT energy consumption and can enable global optimization, has so far received less attention. An exception is the work by Zilberman et al.~\cite{Zilberman:Carbon:2022}, who identify carbon-aware networking as a high-potential research area and sketch the concept of ``carbon-intelligent routing'', i.e., to leverage differences in network paths' carbon intensity (i.e., carbon emission per unit of data transmitted) to reduce the carbon footprint of communications.

Previous research on green inter-domain networking applies carbon efficiency to the optimization metric of Border Gateway Protocol (BGP)~\cite{NAFARIEH201325}. Unfortunately, this direction faces several challenges. \textbf{Inefficient Green Route}: A strict carbon-optimal path can result in a highly inefficient end-to-end path in terms of monetary cost, latency, bandwidth, loss, or jitter (cf.~\Cref{sec:path_quality}). Depending on the application requirements, an optimization subject to all these constraints needs to be made, requiring path selection within a fine-grained metric space. \textbf{Ossification}: Carbon-optimal paths can thus only be offered as additional options, not as replacements for the conventional BGP route. When using BGP to provide carbon-efficient alternative paths, routers would thus require multiple forwarding tables, and packets would need to indicate the desired optimization criteria. Updating BGP and router hardware represents a challenge---as we have experienced in securing the BGP protocol through BGPSEC~\cite{rfc8205}, which has been an effort for over two decades. \textbf{Volatility of Carbon Emissions}: Renewable energy resources (e.g., solar and wind) can fluctuate greatly within short time spans. The dynamic nature of carbon data used in routing metrics would introduce a large number of re-routing events, likely pushing BGP beyond its scalability limits.

This paper aims to overcome these challenges to enable carbon-intelligent inter-domain routing. Concretely, we pose two main research questions in this paper: 
\begin{enumerate}
	\item ~\emph{How can we design a viable system that enables carbon-intelligent inter-domain routing?}
	\item ~\emph{How beneficial for endpoints can this system be in terms of the carbon footprint of their Internet usage?}
\end{enumerate}

To answer the first question, we identify a promising opportunity in Path-Aware Networking (PAN). PAN architectures provide endpoints with path information and allow them to choose paths based on their criteria. Thus, endpoints can optimize paths for multiple criteria, potentially solving the inefficient green route problem. Furthermore, some PANs employ the packet-carried forwarding state where packets convey the forwarding path information in the header field, making routers stateless and addressing the ossification problem. 

\begin{figure}[!t]
	\includegraphics[width=\linewidth]{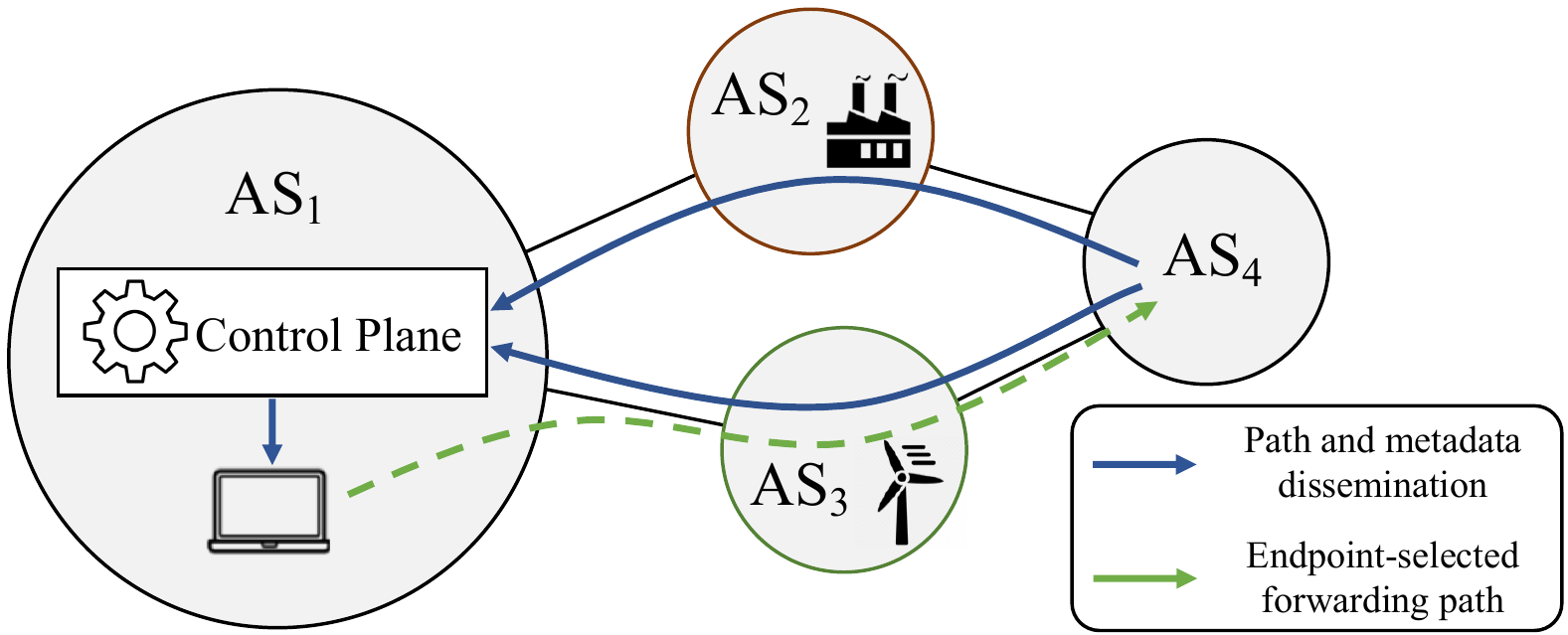}
	\caption{The concept of carbon-intelligent inter-domain routing in a PAN architecture.}
	\label{fig:concept}
\end{figure}

One way to achieve carbon-intelligent routing in PANs is to provide endpoints with paths' carbon information and let them perform a fine-grained path optimization, weighing carbon intensity against other performance criteria~\cite{draft-cx-green-metrics}.~\Cref{fig:concept} illustrates carbon-intelligent inter-domain routing in a PAN architecture. To that end, a PAN architecture needs to be extended to provide carbon information in a practical way. More precisely, path carbon intensity needs to be estimated and disseminated in a way that (1)~provides up-to-date information, (2)~is scalable despite the volatility of carbon intensity, and (3) respects autonomous systems' (AS\footnote{In this paper, we use AS and domain interchangeably.}) privacy concerning their topology and electricity providers. We discuss these requirements in more detail in~\Cref{sec:requirements}, and~\Cref{sec:discuss_privacy}.

In this paper, we design~\systemname{} (\textbf{C}arbon-intelligent \textbf{I}nter-domain \textbf{Ro}uting), a system that forecasts and disseminates carbon intensity of inter-domain paths in a PAN architecture. It satisfies all the aforementioned requirements by \emph{forecasting} day-ahead carbon intensity of inter-domain paths in a \emph{distributed} manner and disseminating them (cf.~\Cref{sec:overview} and~\Cref{sec:detailed_design}). To compute these forecasts,~\systemname{} uses our new model for the carbon intensity of inter-domain paths (cf.~\Cref{sec:model}). 

To prove that scalable green routing can be deployed and used in the short term, we build~\systemname{} based on the commercially deployed SCION PAN~\cite{scion2021scion,Krahenbuhl2021deployment} architecture, implement a proof of concept for it on the open-source SCION codebase~\cite{scionproto} and operate it on the SCIONLab~\cite{Kwon:2020:SCIONLab} global testbed. We anticipate that a concrete open-source system provides an important baseline for future research to extend and improve upon. 

Further, to show the carbon-footprint benefits for endpoints and end domains,
we carry out an investigation by simulating~\systemname{} on a large-scale realistic Internet topology. This topology reflects environmentally relevant characteristics of today's Internet and allows for the simulation of carbon-aware path selection. Our simulations show that~\systemname{} can reduce the carbon intensity of communications by at least 47\% for half of the domain pairs and the carbon footprint of Internet usage by at least 50\% for 87\% of end domains.

In summary, this paper makes the following contributions:
\begin{itemize}
	\item Propose a carbon-intensity model for inter-domain paths;
	\item Design~\systemname{}, a system to forecast the carbon intensity of inter-domain paths and disseminate it to endpoints;
	\item Implement a proof of concept for~\systemname{} on SCION's open-source codebase and operate it on the SCIONLab testbed;
	\item Investigate the impact of car\-bon-aware inter-domain path selection on the carbon footprint of endpoints and end domains by implementing~\systemname{} in the ns-3-based SCION simulator and simulating it on a large-scale topology.
\end{itemize}

%% file: background.tex
\section{Background}
\label{sec:background}

\systemname{} is designed and implemented based on the SCION PAN architecture. Therefore, we provide an overview of SCION. Furthermore, to compute the carbon-intensity forecast of network paths, we employ power-grid carbon-intensity forecasting tools, which we briefly present.

\subsection{PAN Architectures and SCION}\label{sec:scion}
PAN architectures provide endpoints with information about different network paths and let them select paths according to provided information. Thus, PAN architectures introduce a promising opportunity for carbon-aware path selection. SCION is an example of PAN architectures that can help concretize the principles of carbon-intelligent inter-domain networking throughout this paper. Here, we present the SCION features that are directly relevant for the remainder of the paper; a comprehensive description of the architecture can be found in dedicated works~\cite{scion2021scion}.

\paragraph{Architecture} SCION groups ASes in \emph{Isolation Domains} (ISDs). In each~ISD, \emph{core ASes} provide connectivity to other ISDs. Other ASes in an ISD are direct or indirect customers of core ASes.

\paragraph{Data Plane} SCION uses packet-carried forwarding states to forward inter-domain traffic. This means that endpoints encode AS-level inter-domain paths into packet headers, based on which border routers of every AS forward packets. SCION paths are specified at the granularity of ingress and egress interfaces representing links between neighboring ASes, providing endpoints with fine-grained control over paths.

Thus, border routers do not need to store inter-domain forwarding tables to make forwarding decisions, enabling scalable multi-path forwarding and allowing path optimization across multiple criteria, including carbon intensity. 

\paragraph{Control Plane} SCION constructs and disseminates inter-domain paths using a hierarchical control plane with two levels: (1) among all core ASes, and (2) within every ISD along provider-customer links. Each level of the hierarchy is responsible for constructing path segments and disseminating them to endpoints, which combine these segments into complete end-to-end forwarding paths.

Path segments are constructed through the \emph{beaconing} process, a collaborative hop-by-hop process among \emph{beacon service}s of participating ASes that involves exchanging special routing messages called Path Construction Beacons (PCBs). The two hierarchy levels of beaconing are: (1) \emph{core beaconing} among all core ASes in all ISDs, constructing core-path segments, and (2)~\emph{intra-ISD beaconing} within each ISD along provider-customer links, constructing up- and down-path segments.
During the beaconing process, each AS encodes information about its AS hop in an \asentry{} in the PCB. An \asentry{} can contain a \staticextension{} field~\cite{anapaya_docs}, providing path metadata. 

Beacon services extract path segments from PCBs and register them to \emph{path service}s, responsible for resolving endpoints' queries for path segments.
To construct an end-to-end forwarding path, an endpoint retrieves path segments from path services and selects a desirable combination of up, core, and down segments based on their metadata.

\subsection{Forecasting Power-Grid Carbon Intensity}\label{sec:background:forecasting}
The carbon intensity of network paths depends on the carbon intensity of the electricity ($\mathit{CIE}$, i.e., \co{} mass emitted per unit of generated electrical energy in \SI{}{\gram\per{\kilo\watt\hour}}) that network devices consume. Therefore, to forecast the carbon intensity of network paths, we need $\mathit{CIE}$ forecasts, which is an emerging research area. Commercial tools such as ElectricityMaps~\cite{electricitymaps:2022} and WattTime~\cite{watttime:2022} and research projects such as DACF~\cite{Maji:2022:DACF} are example systems that provide short-term $\mathit{CIE}$ forecasting. These tools provide day-ahead hourly forecasts of $\mathit{CIE}$, i.e., a sliding window of forecasted values with one-hour update frequency.

%% file: principles.tex
\section{Design Principles}
\label{sec:principles}
This section describes the design requirements and challenges of a system that estimates and disseminates carbon-intensity of inter-domain paths in a path-aware Internet and provides an overview of~\systemname{}, which overcomes these challenges.

\subsection{Requirements and Challenges}
\label{sec:requirements}
We identify three requirements for estimating and disseminating network paths' carbon intensity across the Internet.

\textbf{R1: Reliable Carbon Information.} The carbon intensity of network paths must be quantified in an accurate and reliable manner. This characterization is complex because the carbon intensity of a network path varies over time and depends on the location of all devices on the path, and their electricity mix, which in turn depends on the daytime and the weather conditions. 


\textbf{R2: Scalable Dissemination.} Car\-bon-in\-ten\-si\-ty information must be disseminated in a scalable way. This is challenging as paths' carbon intensity fluctuates and needs to be disseminated frequently. However, frequent dissemination of information about numerous Internet paths can overwhelm participating ASes in terms of communication and computation costs.

\textbf{R3: Privacy-Preserving Operation.} The private information of ASes, such as their internal topology and electricity providers, must be kept private because ASes may not be willing to disclose this information. On the other hand, accurate computation of an inter-domain path's carbon intensity requires fine-grained information about all devices on the path across multiple ASes. 

\subsection{System Overview}\label{sec:overview}
We design~\systemname{} such that it satisfies requirements R1--R3 above.
We satisfy the reliability requirement (R1) by proposing a model for the carbon intensity of inter-domain paths and also through forecasting. The scalability requirement (R2) is met by providing forecasts instead of real-time information to avoid frequent re-dissemination upon changes. Lastly,~\systemname{} meets the privacy requirement (R3) through a distributed design with instances in all participating ASes: Each AS deploys a local instance of~\systemname{} that forecasts and disseminates the carbon intensity of intra-domain paths without disclosing detailed information about the AS's topology. The carbon intensity of an \emph{inter}-domain path is then calculated by accumulating the carbon intensities of all its constituent \emph{intra}-domain segments, each traversing one AS (cf.~\Cref{sec:model}).

\begin{figure}[!t]
	\includegraphics[width=\linewidth]{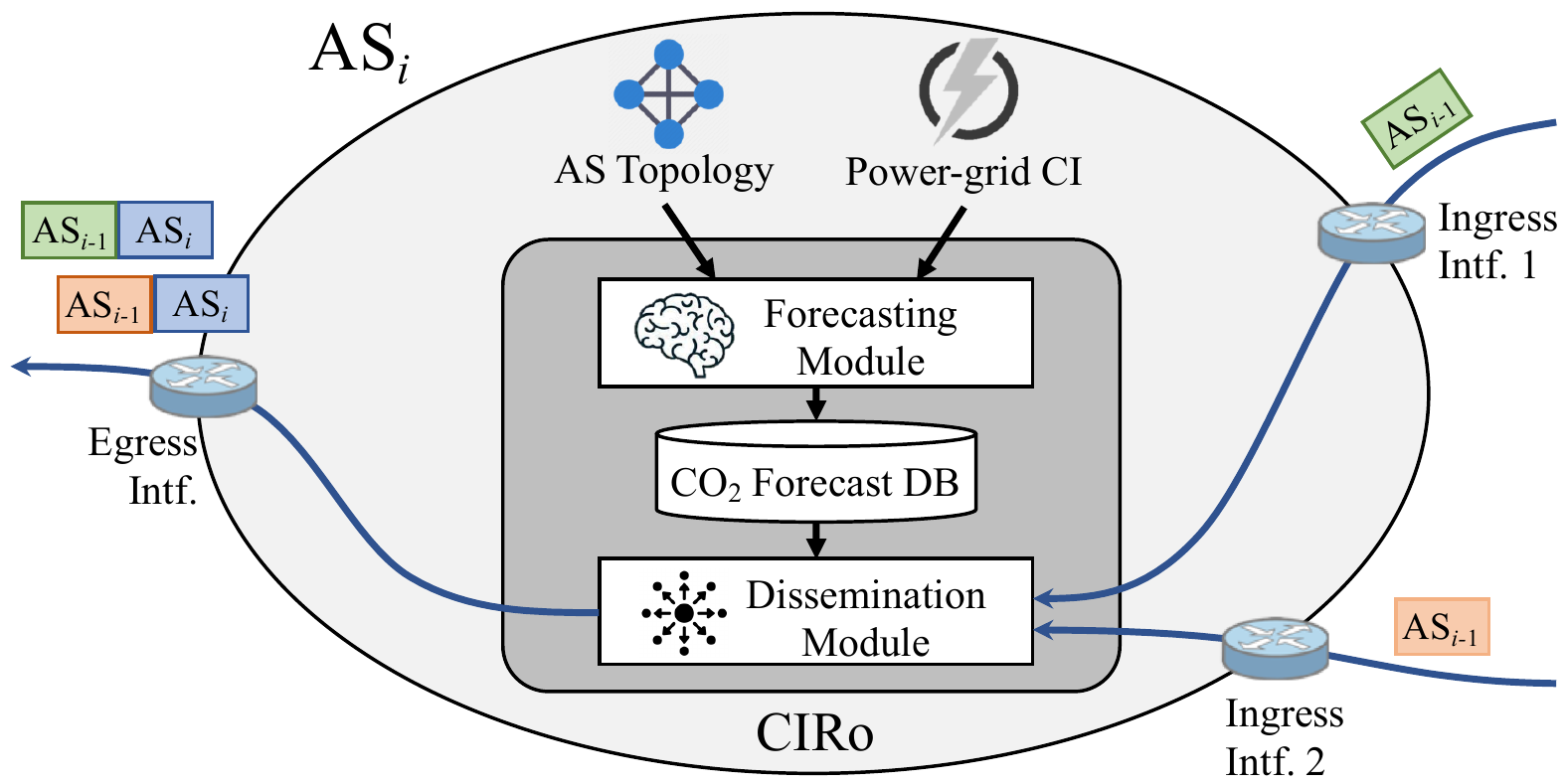}
	\caption{Overview of~\systemname{}. A participating AS runs a~\systemname{} instance. An instance consists of a forecasting module and an information dissemination module. The forecasting module computes path carbon intensity forecasts using a model we introduce, the topology of the AS, and power-grid carbon-intensity (CI) forecasts, and inserts its forecasts into the forecast database. The dissemination module disseminates forecasts to other ASes using routing messages. It also provides endpoints within an AS with these forecasts.}
		\label{fig:overview}
\end{figure}

Each~\systemname{} instance consists of two modules: (1) the forecasting module, and (2) the information dissemination module. The interaction of these modules is illustrated in~\Cref{fig:overview}.

The \textbf{forecasting module} computes day-ahead hourly forecasts (i.e., forecast for 24 hours) for the carbon intensity of intra-domain paths that connect interfaces of the AS. These forecasts are then inserted into the forecast database.
The main building block of the forecasting module is the model we propose for the carbon intensity of inter-domain paths (\textit{CIDT}, i.e., \co{} mass emitted per unit of traffic being forwarded on a path in \SI{}{\gram\per\bit}, cf.~\Cref{sec:model}). This model takes into account the energy intensity of network devices (\textit{EIDT}, i.e., their energy consumption per bit of forwarded traffic in \SI{}{{\kilo\watt\hour}\per\bit}) and the carbon intensity of their used electricity (\textit{CIE}, cf.~\Cref{sec:background:forecasting}). 

The \textbf{information dissemination module} disseminates the carbon intensity forecasts across the Internet. We design this module as an extension to the control-plane infrastructure of a path-aware Internet architecture. We extend the functionality of the control plane to encode carbon-intensity forecasts within routing messages. As a result, endpoints receive this information along with paths without any additional query. Hence, we minimize standardization efforts as well as communication and computation overhead.

%% file: model.tex
\section{Modeling Carbon Intensity of Inter-Domain Paths}\label{sec:model}
In this section, we develop a model enabling~\systemname{} to compute and disseminate carbon-intensity forecasts of inter-domain paths in a distributed manner that respects ISPs' privacy.


\subsection{Carbon Intensity of Data Transmission}
Data transmission over a network path may cause~\co{} emission through two levels of indirection: forwarding traffic causes (electrical) energy consumption on network devices, and the electricity used by each device is generated from energy resources in a process that may emit~\co{}, depending on the energy resources.

We define the \textbf{C}arbon \textbf{I}ntensity of \textbf{D}ata \textbf{T}ransmission ($\textit{\cidt{}}$ in \SI{}{\gram\per\bit}) over a network path as the \co{} emission that can be attributed to a unit of data for being transmitted over that path. Since energy is an additive quantity, the \co{} emission of energy consumption and the \co{} emission of data transmission are additive.
\begin{sloppypar}

\subsection{\textbf{Inter-Domain~\cidt{}}} Using~\cidt{}'s additive property, we write the~\cidt{} of an inter-domain path (denoted by $\fullpath{}$) as the sum of~\cidt{}s of consecutive AS hops it consists of:
\begin{align}
\begin{split}
\label{eq:cidt_inter_domain_path}
        \textit{\cidt{}}_{\fullpath{}} = 
        \sum_{\ashopi{} \in \fullpath{}} \textit{\cidt{}}_{\ashopi{}},
\end{split}	        
\end{align}
where $\ashop{}$ denotes an AS hop on the inter-domain path. $\ifpair{}$ denotes the ingress and egress interface pair from/to which the inter-domain path enters/exits the AS hop.
\end{sloppypar}

The additive property is essential for distributed functionality of~\systemname{}:  each~\systemname{} instance only needs to compute the~\cidt{} of paths within one AS without disclosing topology information.

\Cref{eq:cidt_inter_domain_path} does not explicitly include inter-domain \emph{links} connecting the border routers of neighboring ASes. That is because such links are either (1) direct links connecting two routers in the same data center, or (2) peerings via IXPs, IXP federations, IXP port resellers, or layer-2 links to an IXP. In the first case, the~\cidt{} of the link is included in the~\cidt{} of its neighboring ASes. In the second case, previous research suggests IXPs and resellers constitute ASes in a PAN context~\cite{Krahenbuhl2021deployment}. Thus, they compute their~\cidt{} as normal ASes in the same way we propose in this section.

\subsection{\textbf{Per-Hop~\cidt{}}} An AS may forward traffic on multiple internal paths, e.g., using equal-cost multi-path (ECMP). Therefore, we define the~\cidt{} of $\ashop{}$ as the mean~\cidt{} of all active \emph{intra-domain} paths from its ingress interface to its egress interface. Thus,
\begingroup
\begin{align}
\begin{split}
\label{eq:cidt_each_as}
	 \textit{\cidt{}}_{\ashop{}} = \frac{1}{|\mathcal{P}_{\ifpair{}}|} \sum_{P_{\ifpair{}} \in \mathcal{P}_{\ifpair{}}} \textit{\cidt{}}_{P_{\ifpair{}}}
\end{split}	 
\end{align}
\endgroup
where $P_{\ifpair{}}$ is an \emph{intra-domain} path connecting \textit{ing} and   \textit{eg} interfaces within the AS, and $\mathcal{P}_{\ifpair{}}$ is the set of all such paths. In case of weighted ECMP (WECMP),~\Cref{eq:cidt_each_as} is substituted with a weighted mean over all paths.

\subsection{\textbf{\cidt{} of a Single Intra-Domain Path}}The~\cidt{} over a network path within an AS is composed of two components: (1) the marginal~\cidt{}, which is the result of the actual amount of energy consumed to forward a bit of data over a path, and (2) amortized~\cidt{}, which is the result of energy consumed to keep network paths operational, irrespective of traffic volume over the path. Therefore, 
\begingroup
\begin{align}
\begin{split}
\label{eq:cidt_each_path}
	 \textit{\cidt{}}_{P_{\ifpair{}}} = 
	 \textit{\cidt{}}_{P_{\ifpair{}}}^{\textit{marginal}}+
	 \textit{\cidt{}}_{P_{\ifpair{}}}^{\textit{amortized}}
\end{split}	 
\end{align}
\endgroup
Because of the additive property, we can write
\begingroup
\begin{align}
\label{eq:cidt_from_devices}
        \textit{\cidt{}}_{P_{\ifpair{}}} &=
        \sum_{\substack{\textit{D} \in P_{\ifpair{}}}} \textit{\cidt{}}_{D}^{\textit{marginal}} + \textit{\cidt{}}_{D}^{\textit{amortized}}
\end{align}
\endgroup
where $D$ denotes a network device on $P_{\ifpair{}}$. Since inter-domain communication mostly takes place over backbone networks, we consider typical backbone-network devices in this work, e.g., IP/MPLS core routers and optical wavelength-division multiplexed (WDM) devices~\cite{Energy-AwareDesignofMultilayerCoreNetworks,multilayer}. 
\begin{sloppypar}

\subsubsection{\textbf{Marginal CIDT}} For each device $D$, $\textit{\cidt{}}_{D}^{\textit{marginal}}$ is the result of (1) the marginal energy the device consumes to forward one bit of data (or marginal \textbf{E}nergy \textbf{I}ntensity of \textbf{D}ata \textbf{T}ransmission, $\mathit{EIDT}_{D}^{\textit{marginal}}$ in \SI{}{{\kilo\watt\hour}\per\bit}), and (2) the marginal per-bit energy consumption of the device's external cooling and facilities, which is proportional to $\mathit{EIDT}_{D}^{\textit{marginal}}$~\cite{multilayer}. Thus, the total marginal energy is also proportional  to $\mathit{EIDT}_{D}^{\textit{marginal}}$ by a constant factor ($\eta_{\textit{D},\text{c}}$) determined by the power-usage efficiency (PUE)~\cite{green_grid, multilayer}.

 To compute $\textit{\cidt{}}_{D}^{\textit{marginal}}$, we multiply the total marginal energy by the $CIE$ (in \SI{}{\gram\per{\kilo\watt\hour}} cf.~\Cref{sec:background:forecasting}) in the location of that device ($\textit{CIE}_{\textit{D}}$). Therefore,
\begingroup
\begin{align}
\label{eq:cidt_marginal}
        \textit{\cidt{}}_{D}^{\textit{marginal}} =
\eta_{\textit{D},\text{c}}\mathit{EIDT}_{\textit{D}}^{\textit{marginal}}  \textit{CIE}_{\textit{D}}
\end{align}
\endgroup
\end{sloppypar}


$\mathit{EIDT}_{D}^{\textit{marginal}}$ can be computed using the following formula~\cite{EnergyConsumptionModellingofOpticalNetworks}, irrespective of device type:
\begingroup
\begin{align}
\label{eq:marginal_energy}
        \mathit{EIDT}_D^{\textit{marginal}} = \frac{1}{3600}\frac{P_{D, \mathit{max}} - P_{D, \mathit{idle}}}{C_{D, \mathit{max}}}
\end{align}
\endgroup
where $P_{D, \mathit{max}}$ is the maximum power consumption of the device (in \SI{}{\kilo\watt}), $P_{D, \mathit{idle}}$ is its idle power consumption (in \SI{}{\kilo\watt}), and $C_{D, \mathit{max}}$ is its maximum capacity (in \SI{}{\bps}). Note that \SI{1}{{\kilo\watt\hour}\per\bit} = \SI{3600}{{\kilo\watt}\per\bps}.

\subsubsection{\textbf{Amortized CIDT}} Equivalently to the marginal carbon intensity,
the amortized carbon intensity $\mathit{CIDT}_D^{\mathit{amortized}}$ of a device~$D$ is determined by the amortized idle energy consumption 
$\mathit{EIDT}_{D, \mathit{idle}}^{\textit{amortized}}$,
scaled by the PUE factor~$\eta_{D,c}$. In addition, the model takes into account the set~$\mathcal{R}(D)$ of
\emph{redundant} devices associated with device~$D$, i.e., devices enabled when the primary device~$D$ becomes unavailable.
We assume that these redundant devices do not forward any traffic. 
As a result, we can write:
\begingroup
\begin{align}
\begin{split}
\label{eq:cidt_amortized}
        \textit{\cidt{}}_{D}^{\textit{amortized}} &=
\eta_{\textit{D},\text{c}}\mathit{EIDT}_{\textit{D}, \mathit{idle}}^{\textit{amortized}}  \textit{CIE}_{\textit{D}} \\
&+ \sum_{\substack{\textit{D}^{\prime} \in \mathcal{R}(D)}} \eta_{\textit{D}^{\prime},\text{c}}\mathit{EIDT}_{\textit{D}^{\prime}, \mathit{idle}}^{\textit{amortized}}  \textit{CIE}_{\textit{D}^{\prime}}
\end{split}
\end{align}
\endgroup
Since redundant devices can be located in a different location from the primary device, they could have different $CIE$s.

To amortize the idle energy consumption of a device over a bit of data crossing the device, we hold that bit of data accountable for the device's total \emph{idle} energy consumption during its processing time. The capacity of the device (in \SI{}{\bps}) determines the processing time: On average, processing of a bit on a device takes time $\frac{1}{C_{D, \mathit{max}}}$ (in $\SI{}{\bit}/\SI{}{\bps} = \SI{}{\s}$). By multiplying the idle power consumption of the device by this duration, we arrive at the idle energy consumption of the device during this interval:
\begingroup
\begin{align}
\label{eq:e_idle}
\mathit{EIDT}_{D, \mathit{idle}}^{\textit{amortized}} =\frac{1}{3600} \frac{P_{D, \mathit{idle}}}{C_{D, \mathit{max}}}
\end{align}
\endgroup
In case of redundant devices, we substitute their $P_{D^{\prime}, \mathit{idle}}$ in~\Cref{eq:e_idle}. However, since the \emph{primary} device forwards the traffic, the processing time is still determined by the maximum capacity of the \emph{primary} device. Thus, 
\begingroup
\begin{align}
\label{eq:e_idle_redundant}
\mathit{EIDT}_{D^{\prime}, \mathit{idle}}^{\textit{amortized}} = \frac{1}{3600} \frac{P_{D^{\prime}, \mathit{idle}}}{C_{D, \mathit{max}}}
\end{align}
\endgroup

\subsection{\textbf{Final Form of Per-Hop~\cidt{}}} We combine~\Crefrange{eq:cidt_each_as}{eq:e_idle_redundant} and derive the following formula for the~\cidt{} of a path within an AS between an interface pair:

\begingroup
\begin{align}
\label{eq:cidt_final}
        \textit{\cidt{}}_{\ashop{}} =
        &\frac{1}{3600\cdot|\mathcal{P}_{\ifpair{}}|}
        \sum_{P_{\ifpair{}} \in \mathcal{P}_{\ifpair{}}}
        \sum_{\substack{\textit{D} \in P_{\ifpair{}}}}\\
        &\Bigg(\eta_{\textit{D},\text{c}}\frac{P_{D, \mathit{max}}}{C_{D, \mathit{max}}}\textit{CIE}_{\textit{D}}
        +\sum_{\substack{\textit{D}^{\prime} \in \mathcal{R}(D)}} \eta_{\textit{D}^{\prime},\text{c}}\frac{P_{D^{\prime}, \mathit{idle}}}{C_{D, \mathit{max}}} \textit{CIE}_{\textit{D}^{\prime}}\Bigg) \nonumber
\end{align}
\endgroup

In~\Cref{eq:cidt_final}, the only time-variant variable is $CIE$, which depends on the availability of renewable electricity, which in turn depends on the weather conditions and the time of day. As a result, $\textit{\cidt{}}_{\ashop{}}$ is also time-variant.

%% file: detailed_design.tex
\section{\systemname{}: Design Details}\label{sec:detailed_design}
This section describes the design details of the forecasting and information dissemination modules, for which we implement a proof of concept on the SCION codebase~\cite{scionproto}, available at Anonymous GitHub~\cite{ciro_code}, and run it on the SCIONLab~\cite{Kwon:2020:SCIONLab} testbed. 

\subsection{Forecasting Module}\label{sec:forecast}
The forecasting module computes day-ahead hourly~\cidt{} forecasts between interfaces of each AS. This module is distributed across participating ASes, and each of its instances is administered by one AS. Thus, it does not leak internal information about the AS to any external entity.

\subsubsection{\textbf{Required Inputs}}\label{sec:inputs} Each instance of the forecasting module performs its predictions using~\Cref{eq:cidt_final},
based on the following readily available information from the hosting AS. 

\textbf{Topology Information:}
\begin{enumerate}
	\item \textit{AS Interfaces:} Set of AS interfaces and  the corresponding border routers.
	\item \textit{Intra-Domain Paths:} Device-level intra-domain paths connecting border routers of the AS, computed by intra-domain routing protocols and listed in routers' intra-domain routing information base (RIB).
\end{enumerate}

\textbf{Device Information:}
\begin{enumerate}
	\item \textit{Device Locations:} Locations of all network devices (both primary and redundant) of the AS, including routers and optical devices. If an AS leases optical fibers from other companies, the location and redundancy provisioning of optical devices might not be available. In that case, the module uses models~\cite{multilayer} of IP over WDM networks to approximate locations. Moreover, the same model suggests one redundant device identical to and in the same place as the primary device. 
	\item \textit{Device Specifications:} Maximum power consumption ($P_{\mathit{max}}$), maximum capacity ($C_{\mathit{max}}$), and idle power consumption ($P_{\mathit{idle}}$) of all network devices in the AS network, available in device specifications. If this information is unavailable for device~$D$ (primary or redundant), the module substitutes $\frac{P_{D, \mathit{max}}}{C_{D, \mathit{max}}}$ in~\Cref{eq:cidt_final} with typical values for the energy intensity of network devices~\cite{multilayer} (cf.~\Cref{Tab:energy_intensity} in~\Cref{sec:typical_energy}).
\end{enumerate}

\textbf{Electricity Information:}
\begin{enumerate}
	\item \textit{Power-Usage Efficiency (PUE):} Power-usage efficiency at all AS's points of presence (PoPs), determining $\eta_{\textit{D},\text{c}}$. If unknown, the module uses a typical value of~2~\cite{multilayer}.
	\item \textit{Carbon Intensity of Electricity (CIE):} Day-ahead hourly forecast of $CIE$ at all PoPs from electricity carbon-intensity forecasting sources (cf.~\Cref{sec:background:forecasting}).
\end{enumerate}

\subsubsection{\textbf{Forecasting Procedure}}\label{sec:forecasting_procedure}
A periodic procedure retrieves updated inputs and re-computes day-ahead hourly~\cidt{} forecasts for all interface pairs. The periodic re-com\-pu\-ta\-tion is necessary as (1) $CIE$ forecasts are updated with a period of one hour, and (2) paths between interfaces could change even within these 1-hour intervals. Thus, we consider two re-computation periods: $T_{\mathit{CIE}}$, which is set to 1 hour to retrieve updated $CIE$ forecasts at the beginning of every hour (**:00), and $T_{\mathit{path}}$ to collect updated intra-domain RIBs from routers. $T_{\mathit{path}}$'s value depends on the frequency of path changes in one AS and is set by each AS independently. $T_{\mathit{path}}$ is only relevant if it is less than $T_{\mathit{CIE}}$.

At each $T_{\mathit{CIE}}$ (i.e., beginning of every hour), the module first retrieves all inputs mentioned in~\Cref{sec:inputs}. Then, it computes the day-ahead hourly~\cidt{} forecast for every pair of interfaces on both directions (i.e., both could be ingress or egress interfaces) using~\Cref{eq:cidt_final}. The result for each interface pair in each direction is a vector of 24~\cidt{} values associated with the next 24 hours. Finally, the module inserts the results into the \emph{\cidt{} database} shared with the information dissemination module. 

At each $T_{\mathit{path}}$, the module checks whether intra-domain paths between every interface pair have changed compared to the last $T_{\mathit{path}}$. If so, the module computes~\cidt{} forecasts for the new path and updates the database accordingly.

The database stores two records for each interface pair, one per direction. Each record has 26 columns; the first two columns store the ingress and the egress interface, respectively, and the following 24 columns store~\cidt{} forecasts for the next 24 hours associated with that pair in the direction from the ingress interface to the egress interface.

\subsection{Information Dissemination Module}\label{sec:dissemination}
This module uses the routing infrastructure (control plane) of a PAN architecture to disseminate \cidt{} forecasts. It disseminates forecasts across ASes along with path construction by the means of a routing-message extension we introduce. Endpoints can then receive this information when they request paths from the control plane of their ASes. 

The benefits of building this module based on already existing infrastructure are two-fold: (1) low message complexity and communication overhead as it does not send additional messages dedicated to~\cidt{}, and (2) low design and implementation effort by re-using existing protocols. 

In~\Cref{sec:scion_specific}, we develop a concrete prototype of the design we propose in this section by tuning it to work based on the control plane of the deployed SCION PAN architecture. Using SCION's control plane further improves this module in terms of (1) scalability to large topologies due to SCION's hierarchical routing architecture~\cite{Krahenbuhl2021deployment}, and (2) providing up-to-date carbon information due to the periodic routing in SCION. However, this hierarchical routing only gives CIDT forecasts for path segments, not the entire path. We address this challenge in~\Cref{sec:scion_specific}, which is also applicable to other hierarchical routing architectures.

We now introduce \ciextension{}, an extension to routing messages to disseminate~\cidt{} forecasts. Later, we explain how ASes use this extension to convey the~\cidt{} forecasts of their paths during the routing process.

\subsubsection{\textbf{\ciextension{}}} 
\begin{figure}[t]
	\includegraphics[width=\linewidth]{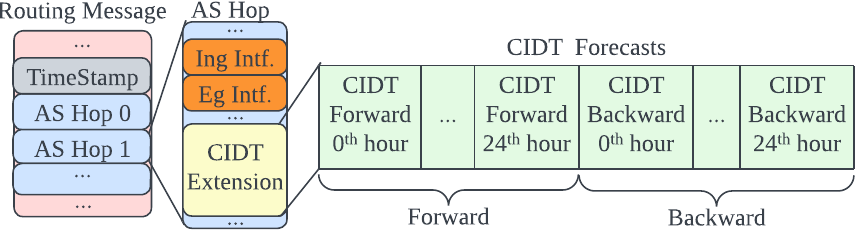}
	\caption{A routing message containing \ciextension{}.}
	\label{fig:cidtextension}
\end{figure}

\Cref{fig:cidtextension} demonstrates the structure of this extension within a routing message. Every AS can include one \ciextension{} in its \texttt{AS Hop} to disseminate its~\cidt{} forecasts. 

A \ciextension{} provides~\cidt{} forecasts for both the forward path from ingress to egress interface ($\textit{\cidt{}}_{\ashopforward{}}$) and the backward path from  egress to ingress interface ($\textit{\cidt{}}_{\ashopbackward{}}$). It is a sequence of 48 contiguous unsigned 8-bit integer numbers, each specifying one~\cidt{} value in \SI{}{\milli\gram\per\giga\bit}. We choose this unit based on our observation of~\cidt{} range in Internet paths (cf.~\Cref{sec:direct_results}). The first 24 values of the vector represent $\textit{\cidt{}}_{\ashopforward{}}$ forecasts for the 24-hour interval starting from the greatest hour before the routing message's \texttt{TimeStamp}. The following 24 values provide similar information for $\textit{\cidt{}}_{\ashopbackward{}}$. This design ensures the time-alignment of forecasts added by different ASes at different times.

For hierarchical routing architectures like SCION, where path segments are combined, \ciextension{} is a map from relative interface identifiers to~\cidt{} forecast vectors, and the structure is different in each routing hierarchy level (cf.~\Cref{sec:scion_specific}).

\subsubsection{\textbf{Timing of~\cidt{} Dissemination}}
Since~\cidt{} forecasts have a 24-hour time window, they need to be disseminated regularly to preserve the temporal continuity of~\cidt{} information. 

This periodic dissemination, however, does \emph{not} require equal frequency or synchronization among ASes as all \ciextension{}s in one routing message are time-aligned.

Therefore, every AS locally chooses its~\cidt{} dissemination period, which is between 1 hour to 2.6 hours. We set the lower bound due to the one-hour update period of~\cidt{} forecasts by the forecasting module, which makes it unnecessary to disseminate~\cidt{} forecasts more than once an hour. We set the upper bound to guarantee the temporal continuity of~\cidt{} forecasts on a path segment in the worst case, where all ASes on the path would synchronously disseminate PCBs with the same frequency. Such a guarantee is satisfied when the routing message received by the last AS contains valid~\cidt{} forecasts for at least one period. This means that this upper bound is $\frac{24}{n-1}$, where $n$ is the number of AS hops on the path segment. Assuming the longest path segment is 10 AS hops long (twice as long as the average AS-hop length in today's Internet~\cite{huston_2021}), the upper bound period is 2.6.

\subsubsection{\textbf{Process of~\cidt{} Dissemination}}
When disseminating \cidt{} forecasts, an AS first selects a set of received routing messages per origin AS (i.e., initiator of the routing message). In this selection, the newest routing messages and the ones with the highest ratio of \texttt{AS Hop}s with~\ciextension{} are prioritized to maximize the temporal and spatial continuity of~\cidt{} information.

Then, for each selected routing message and each egress interface (\texttt{EgIntf}), the AS constructs the corresponding~\ciextension{}, adds it to the \texttt{AS Hop}, extends the routing message with the \texttt{AS Hop} and disseminates it to the egress interface.

To construct the~\ciextension{} for each routing message, the dissemination module first queries the \emph{CIDT database} for~\cidt{} forecasts between \texttt{IngIntf} and the \texttt{EgIntf} in both directions. However, the retrieved results may not be time-aligned with the \texttt{TimeStamp} of the routing message. This is because the database always provides~\cidt{} forecasts for the next 24 hours from the current time, but the \ciextension{} have to provide~\cidt{} forecasts for the next 24 hours after the routing message's \texttt{TimeStamp}, which is in the past. To solve this problem, it shifts the vectors retrieved from the database to the right by the number of hours past from the \texttt{TimeStamp}'s hour. This way, all elements associated with past hours are filled with zeros, and forecasts for beyond 24 hours after \texttt{TimeStamp} are eliminated.

%% file: global_impact/global_impact.tex
\section{\systemname{}'s Impact on Carbon Footprint}
\label{sec:eval}

This section answers the second research question of the paper: To what extent could carbon-aware inter-domain path selection reduce the carbon footprint of the Internet usage of endpoints and end domains (end ASes, i.e., source or destination of the traffic)? This potential is critical to the attractiveness of~\systemname{}, and in turn to the prospect, span, and pace of~\systemname{}'s deployment. We describe our method~\Cref{sec:eval_method}, and we present our results in~\Cref{sec:direct_results}.

\input{global_impact/eval_direct_reductions}

%% file: global_impact/eval_direct_reductions.tex
\subsection{Methodology}\label{sec:eval_method}
Endpoints could benefit from carbon-aware path selection by sending their traffic over paths with lower~\cidt{} (compared to the BGP path in the traditional Internet). This benefit can be found by computing the~\cidt{} difference between \systemname{} and BGP paths and multiplying this difference by the traffic volume between domain pairs.

Since SCIONLab is an overlay network with few virtual ASes, it does not resemble real Internet paths. Therefore, we cannot rely on our proof-of-concept implementation on SCIONLab to approximate real~\cidt{} of paths in the Internet. Thus, we develop a method based on simulations and large-scale realistic datasets. Our method consists of the following steps:
\begin{enumerate}
	\item Approximation of the Internet topology (\cref{sec:eval_method:topo}),
	\item reconstruction of intra-domain paths (\cref{sec:intra-domain-paths}),
	\item carbon-intensity estimation for intra-domain paths (\cref{sec:eval_method:cidt}),
	\item reconstruction of BGP paths for comparison (\cref{sec:eval_method:bgp}),
	\item discovery of \systemname{} paths (\cref{sec:eval_method:greenest}), and
	\item synthesis of a traffic matrix (\cref{sec:eval_method:traffic}).
\end{enumerate}

\subsubsection{\textbf{Creating an Inter-Domain Topology}} \label{sec:eval_method:topo} To find realistic inter-domain paths, we use the inter-domain topology suggested by Kr\"ahenb\"uhl et al.~\cite{Krahenbuhl2021deployment}, which contains a total of 2000 ASes. These are the highest-degree Tier-1 and Tier-2 ASes of the CAIDA AS-Rel-Geo dataset~\cite{CAIDA-Data-Geo}. We extract this topology from the AS-Rel-Geo dataset by iteratively removing the lowest-degree ASes. The benefits of~\systemname{} are most likely realized within these top-tier ASes, as the main differences in the~\cidt{} of paths stem from path diversity provided by massively interconnected ASes in the Internet core.

\subsubsection{\textbf{Finding Intra-Domain Paths}}\label{sec:intra-domain-paths} To compute~\cidt{} of inter-domain paths, we need the~\cidt{} of intra-domain segments they consist of. In turn, the~\cidt{} of intra-domain segments relies on finding intra-domain paths. Since intra-domain paths are not publicly available, we compute them by applying Dijkstra algorithm~\cite{dijkstra1959note} on the internal topology of each AS. 

First, we locate border routers associated with AS interfaces, where the interfaces are available from the AS-Rel-Geo dataset~\cite{CAIDA-Data-Geo}: For each interface between AS~$A_1$ and~$A_2$, we identify all routers of AS~$A_1$ that are also connected to AS~$A_2$, and select the router closest to the interface. This information is available via the CAIDA ITDK dataset~\cite{CAIDA-ITDK}. If no routers satisfy these conditions, we add a router at the interface location. Given these border routers, we compute the shortest router-level path between every pair of border routers by running the Dijkstra algorithm~\cite{dijkstra1959note} on the intra-domain topology of each AS, given by the ITDK dataset. If we cannot find such a path, we assume that the number of routers is an increasing integer-valued step function of the distance between interfaces with the range of $\{1,2,3,4,5\}$ and steps at \SI{1}{\km}, \SI{20}{\km}, \SI{100}{\km}, and \SI{1000}{\km} distance. 

Once router-level paths are established, we add optical devices between consecutive routers. As each packet crosses every WDM switch, transponder, and muxponder once before entering a router and once after leaving it, we assume two of these devices per router on the path~\cite{multilayer}. The number of amplifiers and regenerators, however, depends on the distance between consecutive routers. We assume one amplifier at each \SI{80}{\km} interval, and one regenerator at each \SI{1500}{\km} interval~\cite{multilayer}.

\subsubsection{\textbf{Estimating~\cidt{} of Intra-Domain Paths}}\label{sec:eval_method:cidt}
The~\cidt{} of intra-domain paths is necessary to compute the~\cidt{} of inter-domain paths. Thus, we compute $\textit{\cidt{}}_{\ashop{}}$ for all ingress and egress interface pairs of all ASes in the topology using~\Cref{eq:cidt_final} and computed paths in~\Cref{sec:intra-domain-paths}.

However, not all the required information by~\Cref{eq:cidt_final} is available. First, we do not know using which simultaneous internal paths ASes achieve multi-path routing. Second, we do not know device specifications, the number and location of redundant devices per primary device, and the PUE at PoPs of each AS. Third, hourly $\mathit{CIE}$ (carbon intensity of electricity) forecasts are not available for all device locations.

To address the first issue, we assume single-path intra-domain routing in simulations, as opposed to the real deployment of~\systemname{}. To solve the second problem, we use the model Heddeghem et al.~\cite{multilayer} propose. They suggest typical values for the energy intensity of various types of backbone network devices (cf.~\Cref{Tab:energy_intensity} in~\Cref{sec:typical_energy}). They also propose the typical value of 2 for PUE at each PoP. Further, for each primary device, they consider one identical redundant device. We further assume that both devices are at the same location. To overcome the third challenge, we use each country's \emph{annual average} $\mathit{CIE}$ as the $\mathit{CIE}_D$ of all devices located in that country. We compute the $\mathit{CIE}$ of countries using their mix of electricity-production technology~\cite{iea2018electricity}, and the carbon intensity of electricity-production technologies~\cite{carbonIntensity}. To find the location of routers, we use the CAIDA ITDK dataset~\cite{CAIDA-ITDK}.


\subsubsection{\textbf{Finding BGP Paths and their~\cidt{}}}\label{sec:eval_method:bgp} As not all ASes publish their routing tables, we simulate BGP using the SimBGP simulator~\cite{SimBGP} to find the BGP-selected paths. Then, we compute their~\cidt{} as the sum of~\cidt{}s of all their intra-domain segments. 


\subsubsection{\textbf{Finding Green Inter-Domain Paths}}\label{sec:eval_method:greenest}
To compute green inter-domain paths in the mentioned topology, we implement~\systemname{} in the ns-3-based~\cite{nsnam_ns-3_nodate} SCION simulator~\cite{Krahenbuhl2021deployment,beaconing_simulator}. Furthermore, we change the beaconing (routing) algorithm such that all ASes optimize the~\cidt{} of paths, i.e., they advertise paths with the lowest~\cidt{}. However, as this algorithm is greedy, it may not be able to discover all the greenest available paths.

We use simulations for practicality as setting up a large-scale testbed is not feasible in terms of cost and effort. Further, Krahenb\"uhl et al.~\cite{Krahenbuhl2021deployment} have shown that the SCION simulator is highly predictive of actual testbed results in terms of discoverd paths, so simulation does not affect the reliability of results.

\subsubsection{\textbf{Synthesizing a Traffic Matrix}}\label{sec:eval_method:traffic}
We synthesize a traffic matrix between all ASes using the model proposed by Mikians et al.~\cite{ITMGEN}. We only consider HTTP(S) data and media streaming, and popular video streaming services as they contribute to 60\% of the Internet's traffic~\cite{sandvine}, and can be modeled plausibly. Details are available in~\Cref{sec:traffic_matrix_appendix}.

\subsection{Results}\label{sec:direct_results}
This section presents the result of our study on the effect of carbon-aware inter-domain path selection on the carbon footprint of endpoints and end users. We first compare the~\cidt{} of greenest available paths with BGP paths. Then, we present the absolute and relative reductions in~\cidt{} of communication between each pair of ASes. Finally, we demonstrate the reduction in the carbon footprint of end ASes. 

Note that since the granularity of the traffic matrix is end ASes, not endpoints, we cannot compute the impact on the carbon footprint of individual endpoints. Thus, we only study the impact on their~\cidt{} (i.e., the intensity of carbon footprint), which is equal to the impact on the~\cidt{} of end domains.

\subsubsection{\textbf{Path~\cidt{} Comparison}} 
\begin{figure}[bt!]
\includegraphics[trim = 0cm 0.4cm 0 1cm, width=\linewidth]{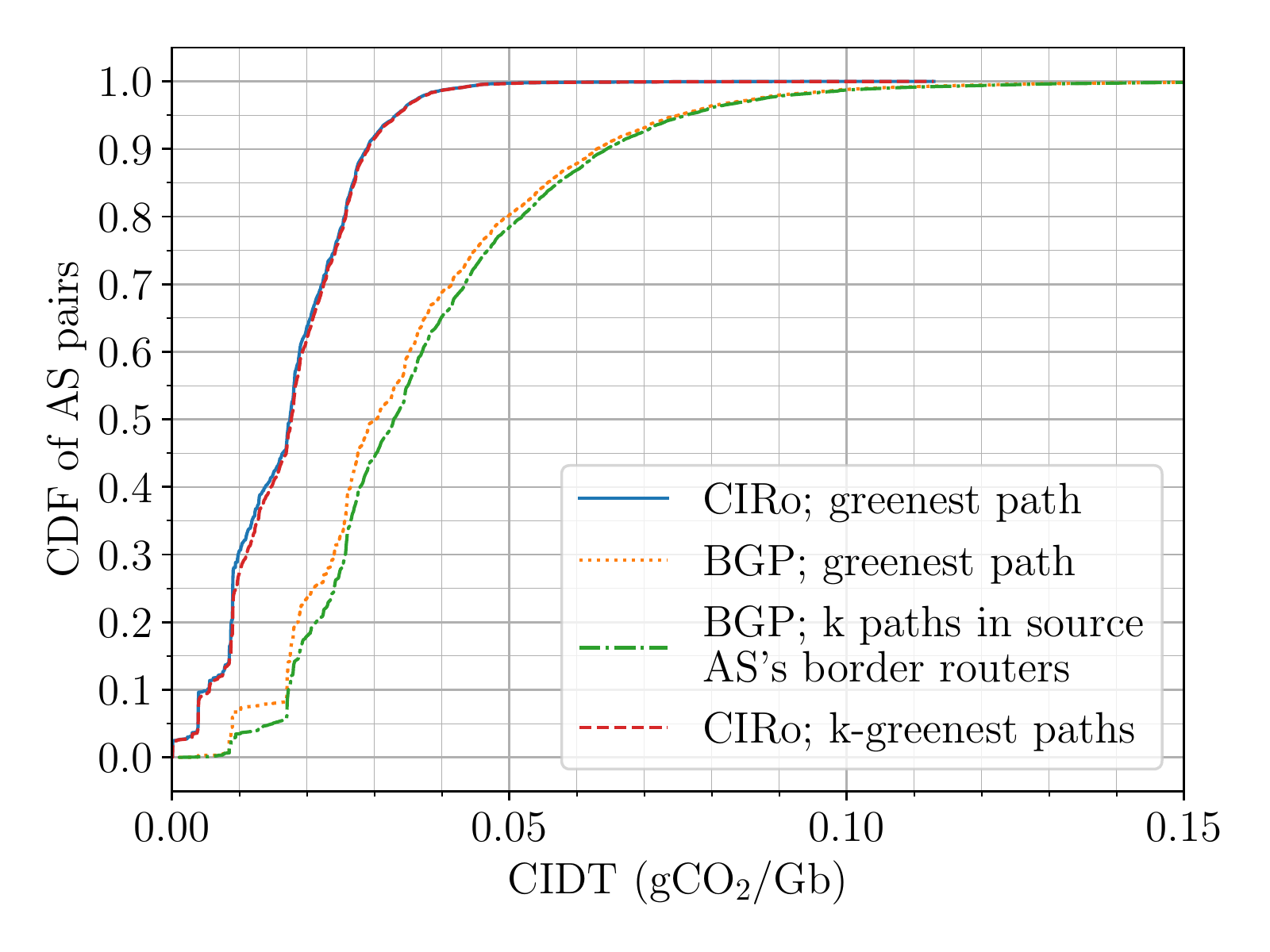}
\caption{Distribution of path \cidt{} across AS pairs, distinguished by path types: (1) the greenest path discovered by~\systemname{}, (2) the greenest BGP path among the $\mathit{k}$ BGP paths in RIBs of the source AS's border routers, $\mathit{k}$ being the number of border routers, (3) the average of available BGP paths, and (4) the average of the $\mathit{k}$-greenest paths discovered by~\systemname{}.}
\label{fig:co2_per_traffic}
\end{figure}

\begin{figure*}[t!]
    \begin{subfigure}[t]{0.3\linewidth}
        \centering
    	\includegraphics[trim =1.2cm 0.4cm 1cm 1cm,  width=\linewidth]{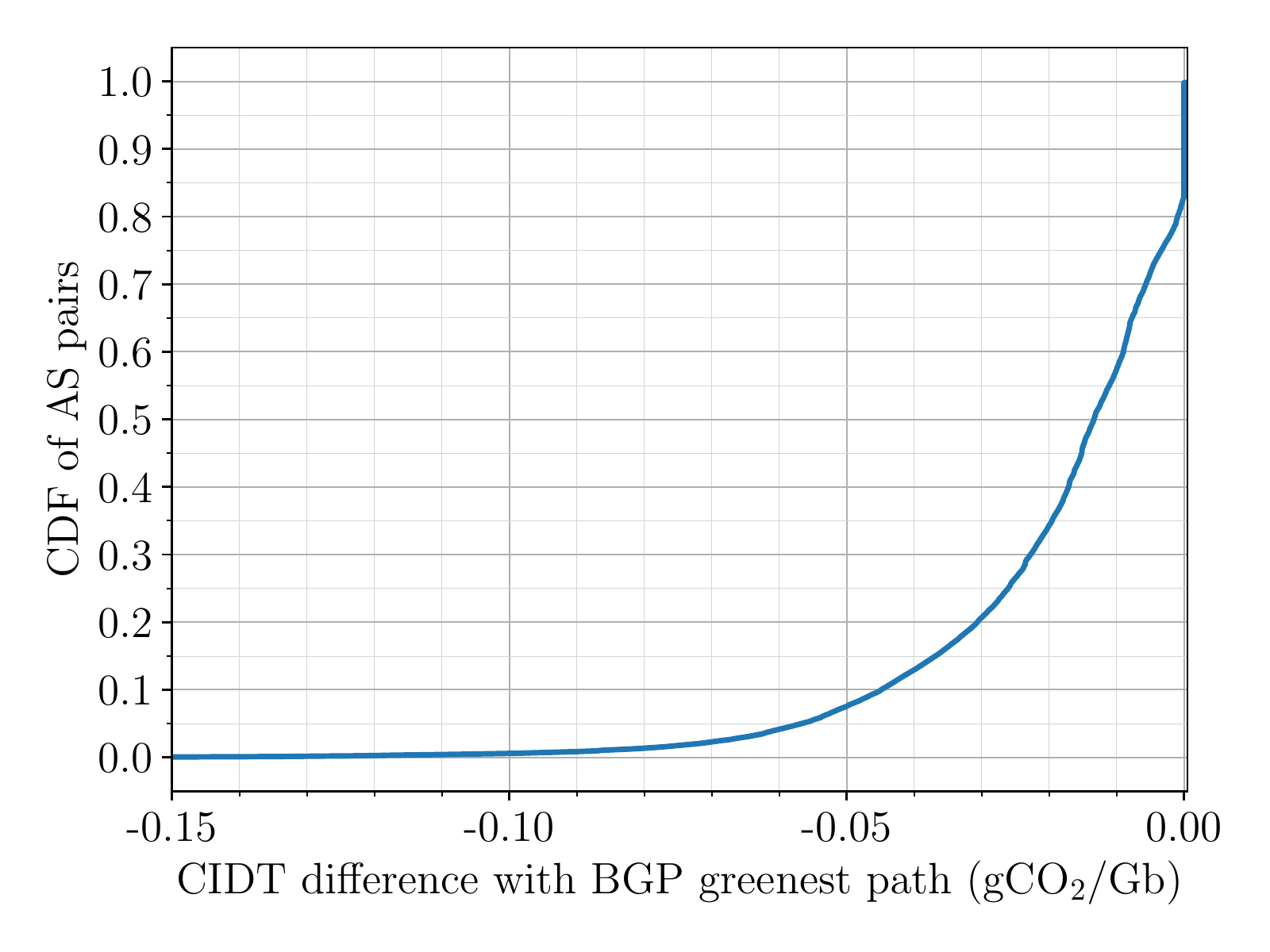}
    	\caption{Absolute~\cidt{} difference between the greenest~\systemname{} path and the greenest BGP path, for all AS pairs.}
    	\label{fig:emission_diff}
    \end{subfigure}\hfill
     \begin{subfigure}[t]{0.3\linewidth}
        \centering
        \includegraphics[trim = 1.2cm 0.4cm 1cm 1cm, width=\linewidth]{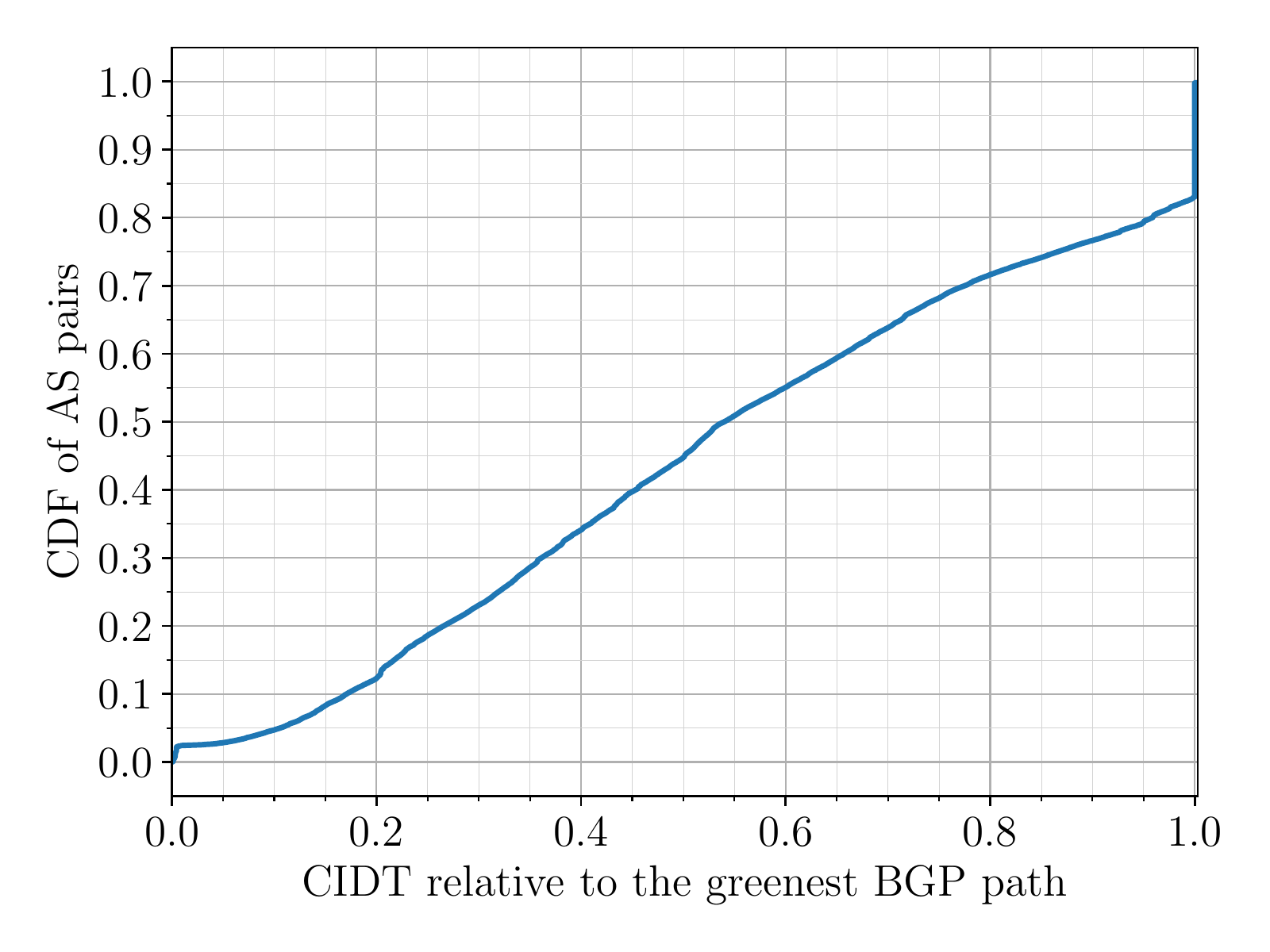}
        \caption{\cidt{} of the greenest~\systemname{} path relative to the greenest BGP path, for all AS pairs.}
        \label{fig:emission_rel}
    \end{subfigure}\hfill
    \begin{subfigure}[t]{0.3\linewidth}
        \centering
        \includegraphics[trim = 1.2cm 0.4cm 1cm 1cm, width=\linewidth]{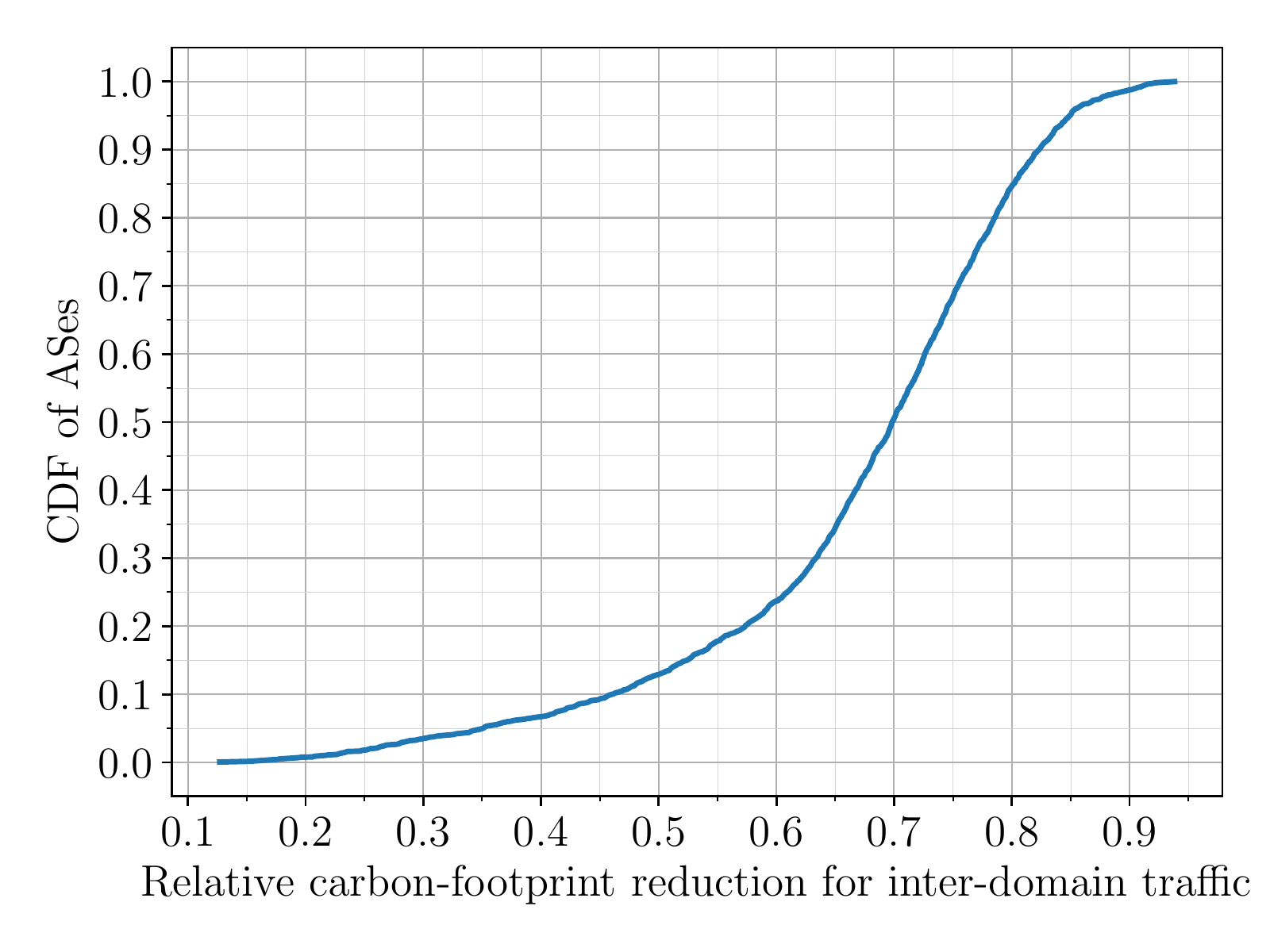}
        \caption{Relative annual reduction in the carbon-footprint of outbound inter-domain traffic, for all ASes.}
        \label{fig:carbon_footprint_savings}
    \end{subfigure}\hfill
    \caption{Greenness comparison between~\copath{}, and BGP.}
    \label{fig:direct_results_figure}
\end{figure*}

\begin{table}[bt!]
\caption{Carbon-footprint reduction for the outbound inter-domain traffic (HTTP(S) and video streaming) of the top-8 beneficiaries of~\copath{} and all ASes.}
    \centering
    \begin{tabular}{lS[table-format=2.0]S[table-format=2.1]S[table-format=6.0]}
        \toprule
        \begin{tabular}{cc} Source AS\\(ASN)\end{tabular} & {\hspace{-13pt}\thead{Carbon footprint\\reduction\\(\SI{}{\ton\per\year})}} & {\thead{Relative carbon\\footprint\\reduction (\%)}} & {\thead{Traffic\\(\SI{}{\exa\byte\per\year})}}\\
        \midrule
        Netflix (2906) & 47640 & 68 & 423\\
        YouTube (36040) & 43032 & 55 & 326\\
        Amazon (16509) & 30240 & 65 & 230\\
        Cloudflare (13335) & 28728 & 79 & 147\\
        Google (15169) & 7536 & 69 & 57\\
        Fastly (54113) & 5520 & 57 & 55\\
        Microsoft (8075) & 4284 & 74 & 29\\
        Incapsula (19551) & 3876 & 76 & 22\\
        \midrule
        All ASes & 210564 & 66 & 2004\\
        \bottomrule
    \end{tabular}
    \vspace*{0.3cm}
    \label{tab:reduction_per_year}
\end{table}

~\Cref{fig:co2_per_traffic} shows the distribution of \cidt{} of the paths available between AS pairs, distinguished by the type of path. It suggests that~\systemname{}-paths almost halve~\cidt{} for almost all percentiles of AS pairs (and their endpoints). It also suggests that~\systemname{} can find \emph{multiple} paths with significantly lower~\cidt{}s than BGP paths.

\subsubsection{\textbf{Absolute \cidt{} Reduction}} \Cref{fig:emission_diff} demonstrates the distribution of \cidt{} difference between the greenest~\systemname{} path and the greenest BGP path, across all AS pairs. It suggests that 83\% of AS pairs (and their endpoints) can benefit from~\cidt{} reduction by using~\systemname{}'s greenest path. This reduction is at least \SI{0.015}{\gram\per\giga\bit} for half of AS pairs. Only a negligible portion of AS pairs experience increased \cidt{} due to the heuristic routing algorithm we use, which may not find the greenest possible path.

\subsubsection{\textbf{Relative~\cidt{}}} \Cref{fig:emission_rel} illustrates the distribution of~\cidt{} of the greenest~\systemname{} path relative to the greenest BGP path, across all AS pairs. According to this figure, half of AS pairs (and their endpoints) can use paths that are at least 47\% greener.

\subsubsection{\textbf{Reduction in Carbon Footprint of End Domains}}\hfill\\
 \Cref{fig:carbon_footprint_savings} depicts the distribution of ASes' relative reduction in annual carbon footprint of their outbound inter-domain traffic enabled by~\systemname{}. According to this figure, this reduction is at least 50\% for more than 85\% of ASes.
 
\Cref{tab:reduction_per_year} demonstrates absolute and relative annual car\-bon-foot\-print reductions through~\copath{} for its top-8 beneficiaries and the sum for all ASes, which is \SI{210}{\kilo\ton}~\co{} or 66\%. It also suggests that the most popular ASes, i.e., the largest CDNs, can benefit from the most \emph{absolute} carbon-footprint reductions.
\vspace*{0.3cm}

%% file: discussion.tex
\section{Discussion}\label{sec:discussion}
This section discusses several important aspects of car\-bon-in\-tel\-li\-gent inter-domain routing and outlines future research directions related to carbon-intelligent routing.

\subsection{Effects on Transmission Quality}
\label{sec:path_quality}

\paragraph{\textbf{Latency}}
\begin{figure}[bt!]
\centering
\includegraphics[trim = 0.2cm 0.4cm 0.2 0.4cm,  width=\linewidth]{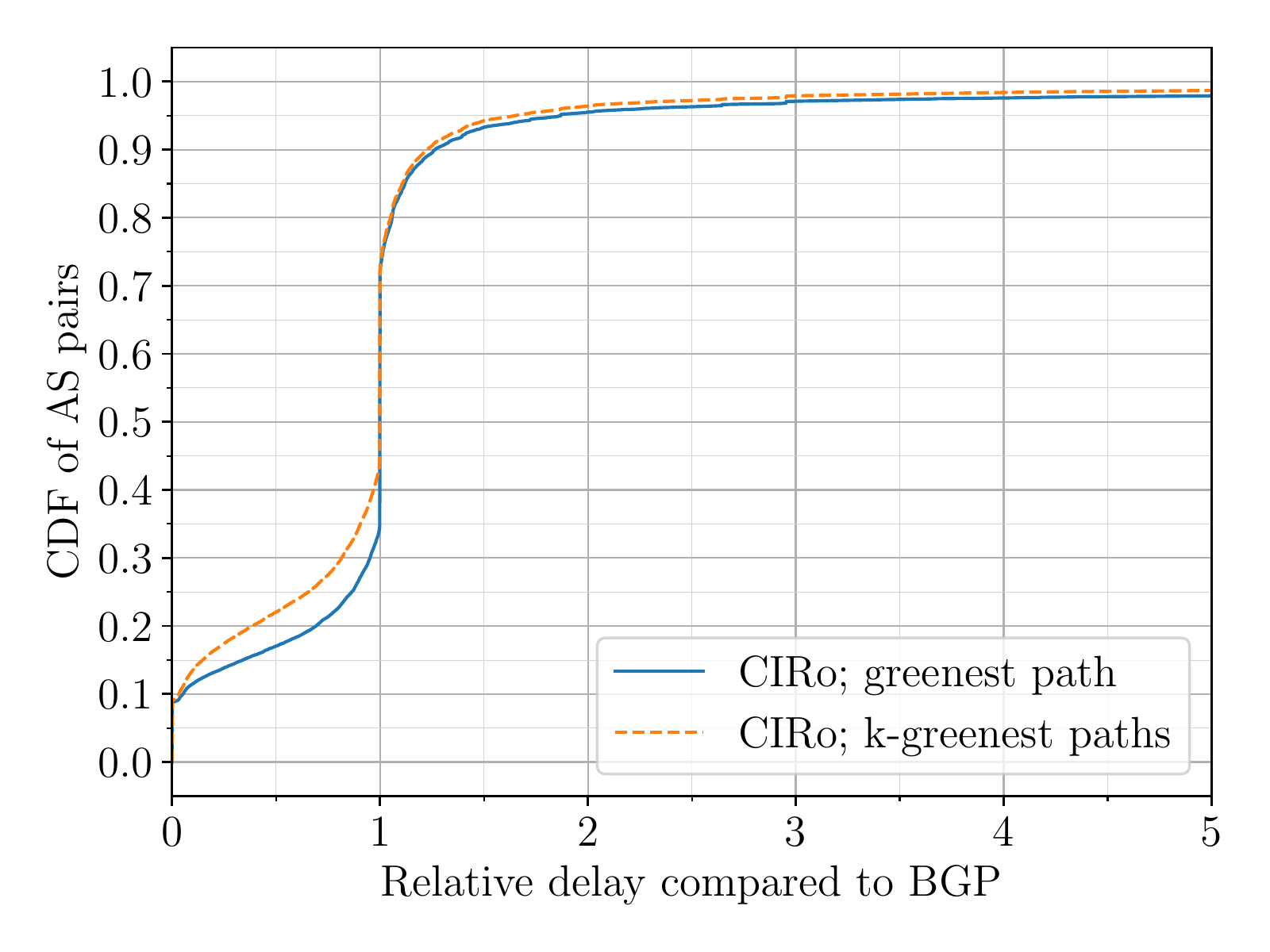}
\caption{AS-pair distribution regarding the relative propagation delay compared to BGP for (1) the greenest path discovered by~\systemname{}, and (2) the average of the $\mathit{k}$-greenest paths discovered by~\systemname{}, $\mathit{k}$ being the number of the source AS's border routers.}
\label{fig:latency_inflation}
\end{figure}
We evaluate the impact of optimizing~\cidt{} of paths on their propagation delay as an indicator of latency. We compute the propagation delay of an inter-domain path as the sum of the propagation delays of all AS hops, which we approximate using the great circle delay between ASes' border routers available in the CAIDA dataset. This is a lower bound for the actual latency, which on average underestimates by a factor of ${\sim}1.5$~\cite{Singla2014}. 

\Cref{fig:latency_inflation} illustrates the distribution of relative propagation delay compared to BGP for paths discovered by~\systemname{}, across AS pairs. According to this figure, optimizing paths only for~\cidt{} does not increase the delay for 75\% of AS pairs, and even reduces it for 35\% of AS pairs. However, around 5\% of AS pairs can suffer from significant inflated delay by a factor of more than 2 due to  \emph{inefficient green routes} resulting from optimizing paths exclusively for~\cidt{}. Nevertheless, end-point path selection in PANs provides an opportunity for multi-criteria path optimization to address this issue. If end-points specify the AS-level path in every packet and border routers can thus be stateless (as in SCION~\cite{scion2021scion}), this multi-criteria optimization can also be scalable. Developing PAN-based multi-criteria path-optimization algorithms is an interesting task for future work.

\paragraph{\textbf{Congestion}} By enabling endpoints to select network paths based on carbon intensity, many endpoints might select green paths, raising concern regarding congestion on such paths. However, due to the vast diversity of paths in the Internet, every AS pair is connected by numerous low-emission paths, with almost the same carbon intensity as the greenest possible paths.~\Cref{fig:co2_per_traffic} confirms that multiple green paths with carbon intensities as low as the greenest path exist between all AS pairs. Thus, endpoints have multiple choices to reduce the carbon footprint of their inter-domain communication, lowering the probability of congestion on one single green path.

\paragraph{\textbf{Traffic Oscillation}} Oscillation arises in load-adaptive path selection, where delayed information about path load entices endpoints to switch paths. This shift causes a high load on the newly selected path and subsequent abandonment of that path~\cite{fischer2009adaptive,scherrer2020incentivizing}. Therefore, such oscillation arises if the path attribute used as a selection criterion changes in response to the selection of the path. In contrast to path load, path carbon intensity is generally no such responsive criterion and therefore is unlikely to cause oscillation. However, we note that path selection based on carbon intensity may be combined with load-adaptive path selection to avoid congestion on green paths. The load-adaptive part of this multi-criteria strategy then needs to follow the guidelines for oscillation-free path selection developed in previous  research~\cite{fischer2009adaptive,kelly2005stability}. We expect congestion risk on green paths to decrease continuously as competitive pressure erodes greenness differences between paths.
\vspace*{0.2cm}

\subsection{Operational Overhead and Carbon Footprint}
\paragraph{Overhead of~\systemname{}}
An important concern is the energy consumption and thus the carbon
footprint of running~\systemname{} itself resulting from its
computation and communication overhead. In the following paragraphs,
we provide a detailed analysis of~\systemname{} overhead, showing that
a single mid-size server is more than sufficient to run a
whole~\systemname{} instance for a large AS with a global
operation. Thus, \systemname{} introduces negligible overhead.

For each instance of~\systemname{}, the overhead of the forecasting module includes \emph{hourly} querying $\mathit{CIE}$ forecasts, collecting AS internal topology and routes, computing~\cidt{} for all interface pairs using~\Cref{eq:cidt_final}, and inserting them into the forecast database. The overhead of the dissemination module includes querying the forecast database, and encoding forecast information into routing messages, increasing the message size.

The overhead of the forecasting module only depends on the size of the AS it is located in. For a large AS with global operation with $O(10^4)$ interfaces, the per-period (hourly) forecasting overhead translates to $O(10^3)$ $\mathit{CIE}$ queries (on the order of number of geographical zones in ElectricityMaps), writing $O(10^8)$ rows in a database table, and $O(10^{10})$ addition and multiplication operations. An efficient implementation using modern databases can carry out this procedure in $O(10^2)$ seconds on a single machine with a few CPU cores with $O(10^1)$ \SI{}{\giga\byte} of memory. We do not consider collecting internal topology and routes as an overhead of~\systemname{}, since ASes already continuously monitor and collect information about their network, irrespective of~\systemname{}'s presence.

The dissemination module's database queries per period (at least once in 2.6 hours. cf.~\Cref{sec:dissemination}) are upper bounded by the square of the number of interfaces in an AS; thus, a large AS with $O(10^4)$ interfaces requires $O(10^8)$ database queries. 

The communication overhead of the dissemination module depends on the number of interfaces of the AS, the number of ASes in the inter-domain topology, and the number of paths per origin AS that the AS advertises to its neighbors. Our SCIONLab experiments with 100 ASes suggest a maximum of \SI{125}{\kilo\byte} overhead per period per interface. By extrapolating to $O(10^4)$ ASes in the network, we arrive at a maximum of \SI{12.5}{\mega\byte} overhead per period per interface. Each inter-domain link needs to transfer this additional data in a period, which is a negligible effort considering the period length and the typical capacity of inter-domain links. 

\paragraph{Overhead of PAN Architectures} Since the stateless routers in a PAN architecture are significantly more energy efficient than routers in today's Internet~\cite{tabaeiaghdaei2022scion_energy}, using a PAN architecture can further reduce the carbon footprint of the Internet.

\subsection{Trustworthy Carbon Information} Green path selection by endpoints could tempt ASes to claim false carbon-intensity information to attract more traffic. Hence, carbon trust roots are needed to certify carbon-intensity information claimed by ISPs. The main challenges for these trust roots are: to estimate the carbon intensity of paths within ISPs, and to ensure that ISPs do not deviate from their certified carbon intensity after obtaining a certificate. 



\subsection{Operator Privacy}
\label{sec:discuss_privacy}
\systemname{} preserves the privacy of ASes by publishing a single carbon intensity number per pair of interfaces instead of disclosing topology and electricity contracts. However, malicious entities might be able to infer useful information about an AS's topology and electricity providers using disseminated information. To address this problem, ASes could augment their~\cidt{} information with randomness to prevent such inferences, or they could encrypt this data and provide trusted entities with the decryption key. The first approach sacrifices data accuracy, while the second approach is still prone to leakage. Thus, fully addressing this issue requires further research.

\subsection{Deployment and Ossification}
The deployment of~\systemname{} in a PAN architecture with stateless routers and a control plane separate from the data plane is relatively easy, as it introduces no change to the data plane and only modest, backward-compatible extensions of the control plane. Thus, no hardware or protocol update is required. 
While the deployment of a PAN Internet itself requires significant effort, the SCION Internet has in fact been experiencing growing deployment~\cite{Krahenbuhl2021deployment} in recent years as a result of its incremental deployability.

\subsection{Economic Effects of Green Routing} From an economic perspective, it is central to predict the impact of carbon routing on the business performance of ISPs and on the competitive environment in the Internet, considering that ISPs might reduce their carbon footprint if they can thereby attract traffic. These competitive dynamics may result in
a virtuous cycle, where ISPs compete for traffic shares by offering paths with ever-decreasing carbon intensity. However, to evaluate the plausibility of this scenario, modeling based on economics and game theory is needed.

\subsection{Green Data Centers and CDNs} A path-aware Internet architecture also enables endpoints to select the data center providing a service or content~\cite{liu2011greening}. Therefore, providing carbon-intensity information of data centers in combination with information about communication paths enables endpoints to monitor and optimize the \emph{total} carbon footprint of their requests, providing \co{} optimization for computation-intensive \emph{and} communication-intensive applications.

%


%% file: related.tex
\section{Related Work}
\label{sec:related}
Much work has studied how to reduce the carbon footprint of the ICT sector by either improving the energy efficiency of end devices and networks~\cite{GreenTE,makingInterdomainPowerAware,GreenGrooming,Jin2017,Sousa2015,GreenIT2012,Wang2017,Jiang2019,Aksanli2012proportionality,NAFARIEH201325}, or increasing the utilization of available renewable energy resources~\cite{EnergyAwareRWA,vanderVeldt2014carbonaware,SWEAR,Raza2014}. 

\paragraph{Green data centers}
A large body of research proposes ecological improvements in the area of data centers~\cite{Gill2018}:
energy-efficient software~\cite{Jin2017}, capacity planning for better resource utilization~\cite{Sousa2015}, 
improved energy management~\cite{GreenIT2012}, better virtualization technologies~\cite{Wang2017}, 
power-saving cooling systems~\cite{Jiang2019}, greenness-oriented load balancing across data centers~\cite{liu2011greening},
or simply using more renewable energy~\cite{Raza2014}.
In a recent work, Radovanovi\'c et al.~\cite{Radovanovic:2022:carbon} propose a method to minimize the carbon footprint of computing among globally distributed data centers by temporally delaying flexible workloads. They achieve this goal by predicting the day-ahead carbon intensity and compution demands.

\begin{sloppypar}
\paragraph{\textbf{Carbon-/energy-aware networking}}
Research conducted on green routing and traffic engineering falls into two main categories: they either make the network more energy-efficient or route packets through paths whose energy resources are green. 

In a visionary paper, Gupta and Singh point out the energy utilization of Internet routers, and suggest energy savings by placing devices in sleep mode~\cite{gupta2003}. 
Later, Zhang et al.~\cite{GreenTE} propose a heuristic method to
reduce energy by turning off line cards and rerouting traffic to 
underutilized links. Vasic et al.~\cite{vasic11} perform an energy 
optimization on real-world networks by inactivating unnecessary
network elements. 
Vasic and Kostic~\cite{vasic2010} propose Energy-Aware Traffic Engineering 
(EATe), to allocate traffic in an energy-optimal manner within an ISP. 
Andrews et al.~\cite{andrews2010} study network optimization from a 
theoretical perspective, with energy minimization as an objective. 
Chabarek et al.~\cite{chabarek08} study the power consumption of core 
and edge routers, optimizing the intra-domain energy consumption. 

Among the studies also taking energy-efficient routing into account, 
Van der Veldt et al.~\cite{vanderVeldt2014carbonaware} propose a path 
provisioning method to reduce the \co{} emission of communications in 
a national research and education network (NREN). 
Aksanli et al.~\cite{Aksanli2012proportionality} design green energy-aware 
routing policies for wide area traffic between data centers.
Ricciardi et al.~\cite{EnergyAwareRWA} use integer linear programming for 
routing and wavelength assignment formulation considering network node's 
carbon intensity. 
Gattulli et al.~\cite{SWEAR} propose a \co{} and energy-aware routing mechanism 
for intra-domain routing. They find two paths for each source and destination pair; one with routers whose energy resources have the lowest emission, and one with the lowest energy consumption. Then, they compare these paths and select the one with the lower emission. 

\end{sloppypar}

\paragraph{Green inter-domain communication}
To the best of our knowledge, only two previous studies propose methods to reduce the carbon footprint of inter-domain communications. Shi et al.~\cite{makingInterdomainPowerAware} extend the aforementioned traffic aggregation method~\cite{GreenTE} by taking into account the traffic between border routers of an AS. Thus, this work does not propose a global inter-domain method as ASes do not cooperate in green routing. Nafarieh et al.~\cite{NAFARIEH201325} propose extensions for OSPF-TE and BGP to propagate information about the emission of links on each path to routers inside and outside an AS. Using this information, each router selects the path with the minimum path emission to all other routers in its own AS, and each border router shares this information with the neighboring AS. During BGP route propagation, the emission of paths is accumulated in update messages, and used by multi-homed ASes to select the provider of the greenest route. However, this BGP-based approach provides limited transparency and control over the carbon footprint of inter-domain communications compared to the present work.

\paragraph{Internet Energy models} Since we propose a model for the carbon-intensity of Internet paths, we provide an overview of the related work in the area of modeling the energy consumption of the Internet in~\Cref{sec:app_related_energy_models} as energy consumption and carbon intensity can be considered as correlated topics.

%% file: conclusion.tex
\section{Conclusion}
\label{sec:conclusion}
In this work, we take a significant step towards realizing carbon-intelligent global routing by designing and implementing~\systemname{}, a practical system to forecast and disseminate carbon-intensity information of inter-domain paths. Our large-scale simulation results suggest that \systemname{} enables endpoints to substantially reduce their carbon footprint of Internet usage.

Importantly, we note that \systemname{} may also support the re\-new\-able-energy transition of entire energy systems
in two main ways.
First, renewable-energy sources such as solar and wind have distinct production peaks, 
which make it challenging to optimally utilize and monetize the electricity produced in these peaks. 
In this regard, \systemname{} is valuable because it allows to shift energy consumption to 
locations where production peaks occur and abundant green energy is currently available.
Second, \systemname{} creates a global competitive environment in which ISPs worldwide can attract
traffic by using green energy. This new competitive pressure is also felt in countries
where the renewable-energy transition has so far been hindered by inert and monopolistic electricity markets.
In these countries, ISPs thus have an incentive to convince local electricity producers
to expand the green-energy supply, which may accelerate the green-energy transition
especially in developing and emerging countries.
In our further research, we are eager to investigate which role exactly carbon-intelligent
routing can play in the global quest for sustainable energy.

%% file: scion_specific_path_combination.tex
\section{SCION-specific details of the Information Dissemination Module}

\begin{figure}[h!]
	\includegraphics[width=\linewidth]{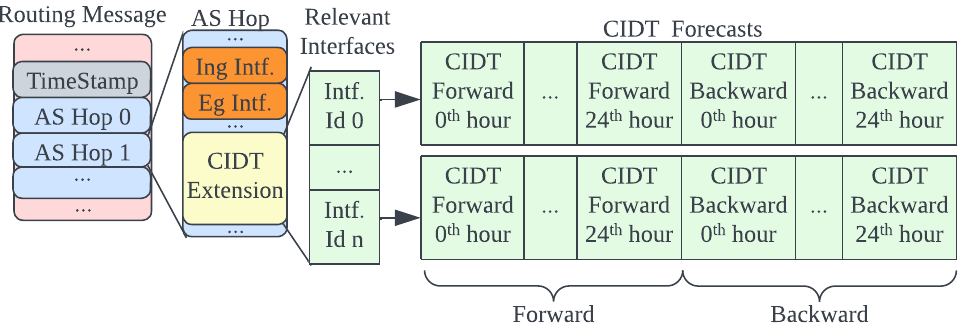}
	\caption{A SCION PCB containing~\ciextension{}.}
	\label{fig:cidtextension_scion_specific}
\end{figure}

\Cref{fig:cidtextension_scion_specific} demonstrates the structure of
this extension within a PCB. Every AS can include one~\ciextension{}
in its~\asentry{} to disseminate its~\cidt{}
forecasts. A~\ciextension{} is a map from relevant interface
identifiers (\texttt{IntfId}) of the AS to vectors of~\cidt{}
forecasts. Having more than one relevant interface is useful for
composition of SCION path segments.
\vspace{5mm}

\label{sec:scion_specific}
\paragraph{SCION-Specific~\ciextension{}}

An \texttt{IntfId}, as a key of this map, represents the intra-domain segment from the \texttt{IntfId} to the egress interface (\texttt{EgIntf}) of the~\asentry{}. 

The~\cidt{} forecast vector associated with an \texttt{IntfId} provides~\cidt{} forecasts for the forward path from \texttt{IntfId} to \texttt{EgIntf} ($\textit{\cidt{}}_{\ashopforward{}}$) and for the backward path from  \texttt{EgIntf} to \texttt{IntfId} ($\textit{\cidt{}}_{\ashopbackward{}}$). 

A~\cidt{} forecast vector is a sequence of 48 contiguous unsigned
8-bit integer numbers, each specifying one~\cidt{} value in
\SI{}{\milli\gram\per\giga\bit}. We choose this unit based on our
observation of the~\cidt{} range in Internet paths
(cf.~\Cref{sec:direct_results}). The first 24 values of the vector
represent $\textit{\cidt{}}_{\ashopforward{}}$ forecasts for the
24-hour interval starting from the full hour time point before the
PCB's \texttt{TimeStamp}. The 24 following values provide similar information for $\textit{\cidt{}}_{\ashopbackward{}}$. This design ensures time-alignment of forecasts added by different ASes at different times.

\paragraph{Database Query by Dissemination Module}
For each PCB to disseminate, the dissemination module queries~\cidt{} forecasts between the \texttt{EgIntf} and all \texttt{IntfId}s in relevant interfaces.

\paragraph{\textbf{CIDT Dissemination in Core Beaconing}}
The forecasting module uses core beaconing to disseminate~\cidt{} forecasts of core-path segments. For each core PCB that is being disseminated by a core AS, the beacon service constructs a \ciextension{} containing the~\cidt{} forecasts only for the intra-domain segment between the ingress and egress interfaces of the PCB. 

Including forecasts for only one interface is essential to the scalability of forecasting module among core ASes. This is because core ASes are large ASes with numerous interfaces, so including information about all interfaces of every hop on the path would introduce significant overhead to each PCB. This per-PCB overhead in conjunction with the high message complexity of core beaconing~\cite{Krahenbuhl2021deployment} would hinder the scalability of the forecasting module. This encoding, however, does not impact the expressiveness of~\cidt{} forecasts since each core PCB only specifies one core-path segment and core-path segments do not combine.

\paragraph{\textbf{CIDT Dissemination in Intra-ISD Beaconing}}
The forecasting module uses intra-ISD beaconing to disseminate~\cidt{} forecasts of up- and down-path segments. Therefore, beacon services include the~\cidt{} forecasts between ingress and egress interfaces of disseminated PCBs in \ciextension{}s. However, since an end-to-end path could be a combination of path segments and peering links, it is essential to provide~\cidt{} forecast of their combination as well to arrive at the end-to-end~\cidt{} forecast. To cover the combination with core-path segments, a core AS originating a non-core PCB includes the~\cidt{} forecasts between the \texttt{EgIntf} of the PCB and all interfaces connecting to all other core ASes. To cover the combination of up- and down-path segments, any AS disseminating a non-core PCB includes the~\cidt{} forecasts between the \texttt{EgIntf} of the PCB and all interfaces connecting to all other customer ASes whose \texttt{IntfId} is smaller than \texttt{EgIntf}. This way, the required information is either encoded in the up segment or in the down segment, halving the overhead. To cover the combination of up- or down-path segments with peering links, any AS disseminating a non-core PCB includes the~\cidt{} forecasts between the \texttt{EgIntf} of the PCB and all interfaces connecting to all other peering ASes.

\paragraph{\textbf{Obtaining the~\cidt{} of Full Inter-Domain Paths}} Once an endpoint retrieves path segments to construct an inter-domain path it computes the~\cidt{} of the full path  for the current hour using \ciextension{}s provided by all segments. Note that different path segments have different origination time, thus, their~\cidt{} forecast vectors are not necessarily aligned. Therefore, endpoints compute the index of current time for each path segment independently. Then, they accumulate the~\cidt{} forecast of all AS hops associated with the current hour to arrive at the~\cidt{} of the end-to-end path.

%% file: traffic_matrix.tex
\section{Traffic Matrix Synthesis}
\label{sec:traffic_matrix_appendix}
We compute the global inter-domain traffic matrix for all ASes in the CAIDA AS-Rel dataset~\cite{CAIDA-Data-AS-rel}  using the model provided by Mikians et al.~\cite{ITMGEN}. In this model, the amount of traffic from AS$_i$ to AS$_j$ is the sum of traffic sent from all users and servers in AS$_i$ to all servers and users in AS$_j$ for all application types:
\begingroup
\begin{align}
\label{eq:11}
\begin{split}
      T_{i, j} = \sum_{A \in \textit{applications}}  m_A\big(S_i p_i^A(j) + d_AS_jp_j^A(i)\big),
\end{split}
\end{align}
\endgroup
where $p_i^A(j)$ is the relative popularity of servers in AS$_j$ for users in AS$_i$ with respect to application~$A$, $S_i$ represents the size of an AS in the number of its IP addresses, $d_A$ denotes the asymmetry factor for application~$A$, which is the ratio of the response flow size to the request flow size for application~$A$, and $m_A$ scales the computed relative traffic for application~$A$ to its absolute real traffic volume.
\begin{sloppypar}
We set $\log_{10}d_{\mathit{http}} = 1$ which is in the range of $(0.4, 1.5)$ for HTTP(S) data and media~\cite{ITMGEN}. Furthermore, we assume the size of each AS to be its number of IPv4 addresses from the CAIDA Pfx2AS dataset~\cite{pfx2as}, except Tier-1 ASes with no users and large CDNs with no users requesting a service. Moreover, $p_i^{\mathit{http}}(j)$ follows the Zipf distribution~\cite{zipf1949zipf} with a slope of 1.2~\cite{ITMGEN}. For each source AS, we thus generate a vector of relative popularities from the Zipf distribution. Then, we assign the largest popularity values to the most popular destination ASes, and the remaining popularity values randomly to other ASes. We define the most popular ASes as the ASes whose hosted websites have the largest accumulated inverse ranks in the Tranco1M dataset~\cite{pochat2021tranco}. We find the website to AS mappings using DNS queries, and the Pfx2AS dataset. Once the matrix is computed, we scale it to the global HTTP(S) traffic, which is estimated to be 82 Exabytes per month~\cite{Statista,sandvine}.
\end{sloppypar}
In addition to HTTP(s) traffic, we also include video traffic in the traffic matrix. For this video traffic, we consider Netflix, YouTube, and Amazon Prime Video, which are responsible for 15, 11.4, and 3.7 percent of total Internet traffic, respectively~\cite{sandvine}. We construct the video traffic matrix by assuming that the amount of traffic any other AS receives from these services is proportional to its number of users in Pfx2AS dataset, and these services receive negligible traffic from other ASes. Finally, we add the video traffic matrix to the HTTP(S) traffic matrix.

To construct the traffic matrix for the core topology of 2000 ASes used in our simulation, we assume that  the amount of traffic between two core ASes is the sum of all traffic in the global traffic matrix between their customer cone ASes in the AS-Rel dataset. If an AS is in the customer cones of multiple core ASes, its inbound and outbound traffic is evenly divided across its core AS providers.

%% file: related_models.tex
\section{Related Work on Estimating the Energy Consumption of the Internet}
\label{sec:app_related_energy_models}

Numerous studies  estimate the energy consumption and energy intensity of the Internet, where the methodologies can be categorized in bottom-up, top-down, and model-based approaches.

\paragraph{Bottom-up approaches} Bottom-up approaches generalize the energy-intensity values obtained through direct measurement of selected network devices. The study conducted by Coroama et al. falls into this category~\cite{coroama2013direct_energy}. In that study, they estimate the energy intensity of the communication between two conference locations in Japan and Switzerland, where the communication path was determined in advance. This work is the only work we are aware of that has estimated the energy intensity of a network \emph{path}, and thus has a relation to our proposed estimation of the \emph{carbon intensity} of inter-domain paths.

\paragraph{Top-down approaches} 
In top-down approaches, researchers divide the total energy consumption of a network by the amount of traffic transited by the network over a particular time period. Studies conducted by Koomey et al.~\cite{koomey2004network_electricity}, Taylor et al.~\cite{taylor2008internet_advertising_energy}, Weber et al.~\cite{weber2010music_delivery_energy}, Lanzisera et al.~\cite{lanzisera_nordman_brown_2011}, and Andrae et al.~\cite{andrae2015trendsto2030} are examples of top-down approaches.

\paragraph{Model-based approaches} Model-based approaches rely on modelling parts of the Internet based on network-design principles, and on energy-consumption information of device vendors in order to find the total energy consumption of specific Internet parts. Baliga et al.~\cite{energy_consumption_of_the_internet, Energy_Consumption_in_Optical_IP_Networks}, Vishwanath et al.~\cite{router_energy_model}, and Hinton et al.~\cite{InternetEnergyModel} propose model-based approaches to estimate the energy consumption of the Internet.

%% file: numbers_tables.tex
\section{Typical Energy Intensity of Network Devices}
\label{sec:typical_energy}

\begin{table}[h!]
\centering
\caption{Energy intensities of typical devices in IP and WDM layers~\cite{multilayer}.}
\label{Tab:energy_intensity}
\begin{tabular}{lS[table-format=5]}
\toprule
 {Device  type} & {\hspace{34pt}\begin{tabular}{cc} Energy intensity\\ ($\textit{I}_{\textit{E}}$ in \SI{}{\watt\per\Gbps}${}={}$\SI{}{\joule\per\giga\bit})\end{tabular}} \\
 \hline
 Core router &  10 \\  
 \hline
 WDM switch (OXC) &  0.05 \\ 
 \hline
 Trans/Mux -ponder & 1.5 \\
 \hline
 Amplifier & 0.03 \\
 \hline
 Regenerator & 3\\
\bottomrule
\end{tabular}
\end{table}

\section{Carbon Intensity of Energy Resources}
\label{sec:energy_resources}

\begin{table}[h!]
\centering
\caption{50$^{\textit{th}}$ percentile \co{} emission of different energy resources. From Edenhofer's IPCC report~\cite{carbonIntensity}.}
\label{Tab:carbon_intensity}
\begin{tabular}{lS[table-format=8.0]}
\toprule
 Energy resource & {\hspace{32pt}Carbon intensity (\SI{}{\gram\co{}\per{\kilo\watt\hour}})}\\
 \midrule
 Coal &  1001\\
 Natural gas &  469\\
 Biomass & 230\\
 Solar & 46\\
 Geothermal & 45\\
 Nuclear & 16\\
 Wind & 12\\
 Hydroelectric & 4 \\
\bottomrule
\end{tabular}
\end{table}